\documentclass[twocolumn,tighten]{aastex631}

\usepackage{graphicx}
\usepackage{amsmath}
\usepackage{amssymb}
\usepackage{color}
\usepackage{mathtools}
\usepackage{ulem}
\usepackage[all]{hypcap} 

\newcommand\msun{\, {M}_\odot}

\DeclareRobustCommand{\VAN}[3]{#2}
\let\VANthebibliography\thebibliography
\def\thebibliography{\DeclareRobustCommand{\VAN}[3]{##3}\VANthebibliography}

\begin{document}

\title{On the Likely Dynamical Origin of GW191109 and of Binary Black Hole Mergers with Negative Effective Spin}

\author[0000-0002-2905-9239]{Rachel C. Zhang}
\affiliation{Department of Physics \& Astronomy, Northwestern University, Evanston, IL 60208, USA}
\affiliation{Center for Interdisciplinary Exploration \& Research in Astrophysics (CIERA), Northwestern University, Evanston, IL 60208, USA}

\author[0000-0002-7330-027X]{Giacomo Fragione}
\affiliation{Department of Physics \& Astronomy, Northwestern University, Evanston, IL 60208, USA}
\affiliation{Center for Interdisciplinary Exploration \& Research in Astrophysics (CIERA), Northwestern University, Evanston, IL 60208, USA}

\author[0000-0001-9879-6884]{Chase Kimball}
\affiliation{Department of Physics \& Astronomy, Northwestern University, Evanston, IL 60208, USA}
\affiliation{Center for Interdisciplinary Exploration \& Research in Astrophysics (CIERA), Northwestern University, Evanston, IL 60208, USA}

\author[0000-0001-9236-5469]{Vicky Kalogera}
\affiliation{Department of Physics \& Astronomy, Northwestern University, Evanston, IL 60208, USA}
\affiliation{Center for Interdisciplinary Exploration \& Research in Astrophysics (CIERA), Northwestern University, Evanston, IL 60208, USA}



\begin{abstract}

With the growing number of binary black hole (BBH) mergers detected by LIGO/Virgo/KAGRA, several systems have become difficult to explain via isolated binary evolution, having components in the pair-instability mass gap, high orbital eccentricities, and/or spin-orbit misalignment. Here, we focus on GW191109\_010717, a BBH merger with component masses of $65^{+11}_{-11}$ and $47^{+15}_{-13}$ $\rm M_{\odot}$, and effective spin $-0.29^{+0.42}_{-0.31}$, which could imply a spin-orbit misalignment of more than $\pi/2$ radians for at least one of its components. Besides its component masses being in the pair-instability mass gap, we show that isolated binary evolution is unlikely to reproduce the proposed spin-orbit misalignment of GW191109 with high confidence. On the other hand, we demonstrate that BBHs dynamically assembled in dense star clusters would naturally reproduce the spin-orbit misalignment and the masses of GW191109, and the rates of GW191109-like events, if at least one of the components were to be a second-generation BH. Finally, we generalize our results to all the events with a measured negative effective spin, arguing that GW200225 also has a likely dynamical origin. 

\end{abstract}

\keywords{Black holes(162) --- Gravitational wave sources(677)}


\section{Introduction} \label{sec:intro}

\label{sec:sec1}
The LIGO/Virgo/KAGRA (LVK) Collaboration has recently released the third Gravitational Wave Transient Catalog \citep{ligo21}, which has brought the number of candidate binary black hole (BBH) mergers to more than $90$ events, transforming our understanding of BHs and gravitational--wave (GW) physics \citep{lvc2020catb,lvc2020catc}. With the upcoming fourth observational run and the next-generation observatories, such as LISA \citep{lisa13}, the Einstein Telescope \citep{maggiore20}, and Cosmic Explorer \citep{reitze19}, the number of GW detections will continue to quickly grow.

Despite the growing population of detected BH mergers, their origin is still highly uncertain. Two main formation channels have been discussed to explain the origin of merging compact objects: isolated binary evolution \citep[e.g.,][]{paczynski76, vandenheuvel76, tutukov93, belczynski02, kalogera07, dominik12, dominik13, postnov14, belczynski16a, belczynski16b, stevenson17, vandenheuvel17, giacobbo18, neijssel19, spera19, bavera21} and dynamical assembly in dense stellar environments \citep[e.g.,][]{portegies00,rodriguez15, banerjee17, banerjee18, banerjee18b, dicarlo19, askar17, fragione18, samsing18b, samsing18c, kremer20, fragione21c}. Sub-channels of these two broad categories include chemically homogeneous evolution of close binaries \citep[e.g.,][]{demink09, demink16, mandel16, marchant16}, hierarchical triple and quadruple systems \citep[e.g.,][]{antonini12, hoang18, fragione19, martinez20,hamers21,martinez22} and formation in disks of active galactic nuclei \citep[e.g.,][]{bartos17, tagawa18, tagawa20}. 

The isolated binary evolution and dynamical channels can be distinguished through several characteristic features. In the former case, merging BBHs may have component masses up to about $45\msun$, as dictated by pair-instability physics \citep[e.g.,][]{heger02, woosley07, farmer19}. Moreover, merging BHs have spins preferentially aligned with the orbital angular momentum (implying a positive effective spin), no residual eccentricity at $10$\,Hz, and mass ratios close to unity \citep{kalogera00, dominik13, samsing18b}. On the other hand, BBH mergers catalyzed by dynamical encounters in dense star clusters have an isotropic orientation of spins relative to the orbital angular momentum (implying a symmetric distribution of the effective spin around zero) and a broader spectrum of eccentricities in the LVK frequency band, but still mass ratios preferentially close to unity \citep{rodriguez16b, rodriguez18, rodriguez19, samsing18b, martinez22}. Importantly, the component masses of merging BHs can exceed the limit imposed by pair-instability physics, if they are the remnant of a hierarchical merger \citep[e.g.,][]{antonini19, fragione20b, mapelli21, fragione22}.  

\begin{figure}
    \centering
    \includegraphics[width=\columnwidth]{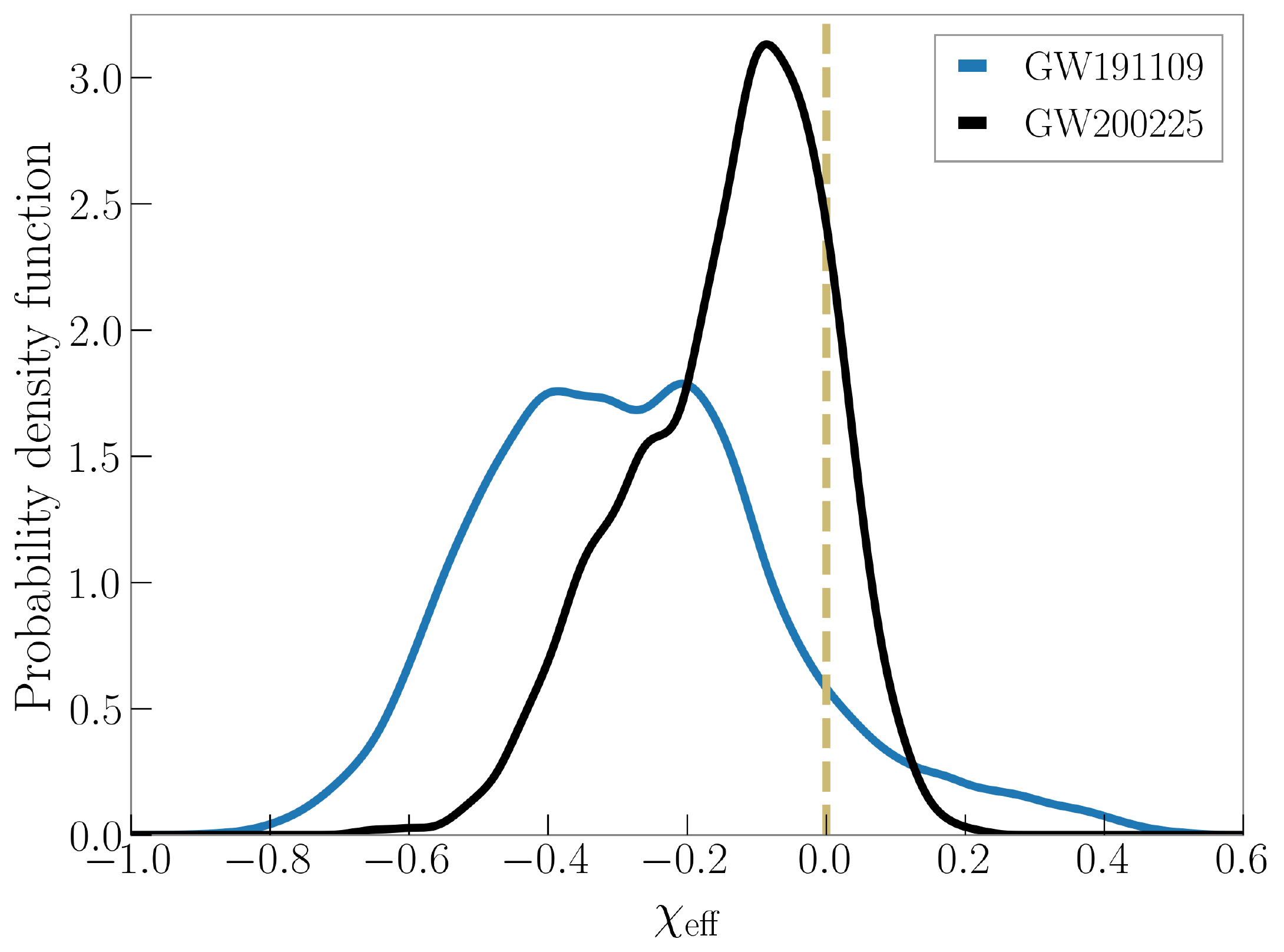}
    \caption{Posterior distributions of the effective spin of GW191109 and GW200225\_060421 derived from the pesummary package by \cite{hoy20}, using the open data from \cite{ligo23}. The probability of a negative effective spin is about 90.6\% and 86.7\% for the GW191109 and GW200225\_060421, respectively.}
    \label{fig:fig1}
\end{figure}

In this paper, we focus on GW191109, the detected BBH merger with 90.6$\%$ of its effective inspiral spin distribution in the negative regime, as shown in Figure \ref{fig:fig1} \footnote{Data origin: \cite{ligo23}; software: \cite{hoy20}.}. Effective inspiral spin is defined as $\chi_{\rm eff}\equiv(m_1\chi_1 + m_2\chi_2)/( m_1+m_2)$, with $m_i$ and $\chi_i$ ($i=1,2$) being the components masses and spins. This BBH merger event has primary and secondary masses of $65^{+11}_{-11}$ and $47^{+15}_{-13}$ $\rm M_{\odot}$, respectively, and effective spin $-0.29^{+0.42}_{-0.31}$, which implies a spin-orbit misalignment of more than $90^\circ$ for at least one of its components. The posterior distribution of the effective spin of GW191109 is shown in Figure \ref{fig:fig1}, along with that of GW$200225\_060421$, the BBH merger event with $86.7\%$ of its effective inspiral spin distribution lying in the negative regime \citep{ligo21}. We argue that the spin orientation is a strong indicator of isolated binary or dynamical formation for GW191109, and in general for any BBH merger with a negative value of its effective spin. We note that the observed effective spin distribution can be affected by glitches, and a glitch was found in the Livingston data for GW191109 \citep{davis22}. Thus, we acknowledge the uncertainties for this measurement and further discuss this in Section \ref{sec:sec4}. We also note that the effective inspiral spin distribution of GW191109 may still be consistent with a $\chi_{\rm eff}\approx 0$ and could be affected by statistical fluctuations and/or model misspecification, which might not entirely rule out the possibility that all events could be explained by isolated binary formation \citep{roulet21, galaudage21, tong22}. However, we note that beyond the effective spin, the measured high masses and inferred merger rate also point towards a likely dynamical origin for GW191109, as explained in Section \ref{sec:sec3}.

Our paper is organized as follows. In Section~\ref{sec:sec2}, we summarize the GW events with characteristic features that indicate an unlikely isolated binaries origin. In Section~\ref{sec:sec3}, we tailor our discussion to GW191109, and show that isolated binary evolution is very unlikely to produce GW191109-like events, while dynamics can easily explain its component masses, effective spin, and merger rates. We conclude and generalize our results in Section~\ref{sec:sec4}.

\section{Binary black holes with an unlikely origin in field binaries}

\label{sec:sec2}

Here, we review BBH mergers which have been discussed to have an unlikely origin as isolated binaries. 

\begin{enumerate}
    \item {\it Masses in the pair instability mass gap range.} BHs with masses approximately in the range $\sim 50\msun-120\msun$ (depending on the progenitor metallicity) are not expected to be formed from stellar collapse, due to runaway pair-instability processes. 
    \citep[e.g.,][]{fowler64, bond84,heger02,woosley07,woosley17, farmer19}. The merger event GW190521 has component masses of about $90\msun$ and $60\msun$, nominally in the pair-instability mass gap. These masses can be naturally produced in dense stellar environments if these BHs are the remnant of a previous merger event \citep[e.g.,][]{miller02b, antonini16, gerosa17, rodriguez19, fragione20c, kimball21}, or in AGN disks through subsequent mergers and gas accretion \citep[e.g.,][]{tagawa20, tagawa20b, tagawa20c}. Other candidate BBH events with at least one component BH in this upper mass gap range include GW$190519\_153544$, GW$190602\_175927$, GW$190706\_222641$, GW$200220\_061928$ \citep{ligo21a, ligo21}.  
    
    \item {\it Unequal mass ratios.} Both isolated binaries and cluster dynamics produce BBH mergers that have preferentially mass ratios close to unity \citep[e.g.,][]{dominik13,belczynski16a,rodriguez16b,rodriguez19}. However, hierarchical mergers in dense star clusters can naturally produce smaller mass ratios \citep[e.g.,][]{fragione22}. GW190412 is a detected BBH merger with a nearly $4:1$ mass ratio, which could easily be explained as a third-generation merger in a massive star cluster \citep{rodriguez20}. Other possibilities include a merger in a hierarchical triple, where the inner binary is driven to a merger by the Lidov-Kozai cycles imposed by the tidal field of the tertiary \citep{su21, martinez22}, or subsequent mergers in AGN disks \citep[e.g.,][]{tagawa20, tagawa20b, tagawa20c}. 

    \item {\it Non-zero orbital eccentricities.} BBHs tend to have zero eccentricity through isolated binary evolution. This is because orbits tend to circularize to minimize energy \citep{peters64}, and merger timescales in isolated binary evolution are long enough for circularization to happen. However, in a dense stellar cluster, highly eccentric BBHs are formed during few-body interactions of BH systems. Specifically, $\sim 5\%$ of all BBH mergers from globular clusters are likely to have an eccentricity $\gtrsim 0.1$ in the LVK frequency band \citep{samsing18b}. Other possibilities to create merger events with large eccentricities is through Lidov-Kozai cycles in hierarchical systems \citep[e.g.,][]{fragione19}. \cite{gayathri22} uses numerical relativity simulations to justify GW190521 as a potential highly eccentric BBH, and \cite{romero-shaw22} argues that GW191109 and GW200208 both have a non-negligible eccentricity in the LVK detection band.

    \item {\it Negative effective inspiral spin parameter.} Isolated binary evolution leads to a preferential alignment between the BH spins and the binary orbital angular momentum, implying typically a positive value of the effective spin \citep{kalogera00}. A positive value of the effective spin may also be preferred by the AGN disk channel if radial migration of BHs is inefficient \citep[e.g.,][]{tagawa20c}. Dynamical assembly in dense star clusters \citep{rodriguez16b} and mergers in hierarchical systems \citep{martinez22} leads to an isotropic distribution of the effective spin around zero. Therefore, a system with a negative value of the effective spin is unlikely to be originated in field binaries and AGN disks (for details see Section \ref{sec:sec3}). Currently, the GW BBH events that show a likely negative observed $\chi_{\rm eff}$ are GW191109 (see Section \ref{sec:sec3}) and GW200225 (see Section \ref{sec:sec4}).

\end{enumerate}

\section{The case of GW191109}
\label{sec:sec3}

\begin{figure*}
    \centering
    \includegraphics[width=0.85\textwidth]{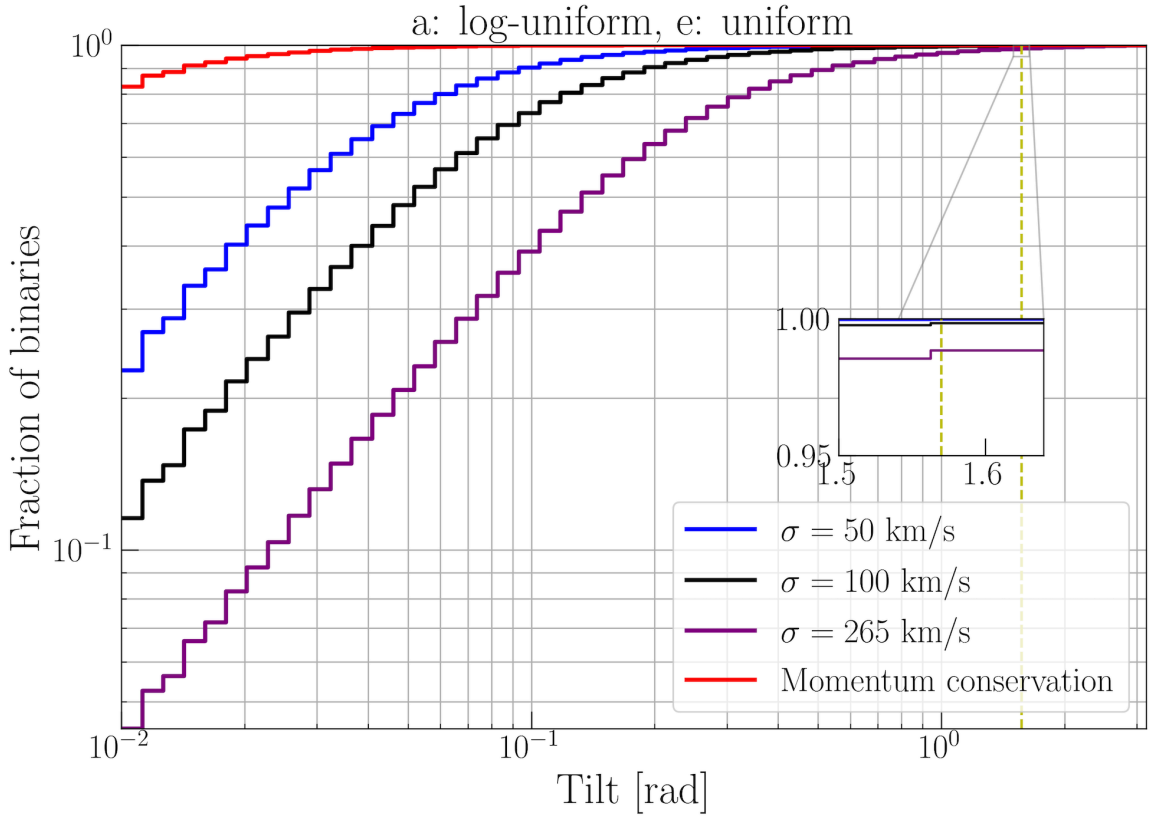}
    \label{fig:fig2}
    \caption{The cumulative distribution of the final tilts for binaries with the following initial conditions: the mass loss distribution is simulated by SSE with initial metallicities of $0.01$ (see Appendix \ref{fig:a2}), the initial semimajor axis distribution is log-uniform with range $10^{-2}$ AU - $10^{-1}$ AU, and the initial eccentricity distribution is uniform. Different colors show different kick velocity distributions: $\sigma = 50$ km s$^{-1}$ (blue), $\sigma = 100$ km s$^{-1}$ (black), $\sigma = 265$ km s$^{-1}$ (purple), and $\sigma = 265$ km s$^{-1}$ scaled by momentum conservation (red). The dotted yellow line marks $\pi/2$, the tilt angle at which the spin of the post-SN mass would be anti-aligned with the orbital angular momentum, possibly producing a negative $\chi_{\rm eff}$.}
\label{fig:2}
\end{figure*}

In this Section, we show that the likely origin of GW191109 is dynamical assembly in dense star clusters. We use the properties masses, effective spin, and merger rate to show that field binaries are unlikely to produce GW191109. Then, we also discuss why other dynamical channels (hierarchical systems and AGN disks) are unlikely to explain this event.

\subsection{Isolated binary evolution}
\label{sec:isolated}
The isolated binary evolution channel for BBH formation involves two massive stars ($\gtrsim 20\msun$) in a relatively close orbit. Typically, the more massive one leaves its main sequence first, expanding in radius and possibly donating mass to the companion when filling its Roche lobe. Eventually, the primary undergoes a supernova (SN) explosion or directly collapses to form a BH. After the secondary also leaves its main sequence and expands, the system can have a phase of either stable mass transfer or unstable mass transfer. In the latter case, a common envelope may form \citep{paczynski76, vandenheuvel76, tutukov93, belczynski02, dominik12, stevenson17, giacobbo18}. After the secondary also undergoes a SN explosion or direct collapse to a BH, a bound BBH is formed, which may merge in Hubble time.

\begin{table*}
    \begin{tabular}{c|cccc}
        \hline
        &
        Momentum conservation
        &
        $\sigma = 50$ km s$^{-1}$
        & 
        $\sigma = 100$ km s$^{-1}$
        &
        $\sigma = 265$ km s$^{-1}$\\
        \hline
         \textbf{Uniform mass loss} \\
        \hline
        a: log-uniform (0.1 AU), e: uniform
        &
        0$^*$
        &
        $5.9 \times 10^{-4}$
        &
        $2.3 \times 10^{-3}$
        &
        $1.5 \times 10^{-2}$ \\ 
        \hline
        a: log-uniform (0.1 AU), e: thermal
        &
        $2 \times 10^{-5}$$^*$
        &
        $1.2 \times 10^{-3}$
        &
        $4.5 \times 10^{-3}$
        &
        $2.7 \times 10^{-2}$ \\ 
        \hline
        a: uniform (0.1 AU), e: uniform
        &
        0$^*$
        &
        $8.4 \times 10^{-4}$
        &
        $3.3 \times 10^{-3}$
        &
        $2.1 \times 10^{-2}$ \\ 
        \hline
        a: uniform (0.1 AU), e: thermal
        &
        $2 \times 10^{-5}$$^*$
        &
        $1.9 \times 10^{-3}$
        &
        $6.6 \times 10^{-3}$
        &
        $3.8 \times 10^{-2}$ \\ 
        \hline
         a: log-uniform (1 AU), e: uniform
        &
        $7 \times 10^{-5}$
        &
        $3.8 \times 10^{-3}$
        &
        $1.2 \times 10^{-2}$
        &
        $4.5 \times 10^{-2}$ \\ 
        \hline
        a: log-uniform (1 AU), e: thermal
        &
        $1.1 \times 10^{-4}$
        &
        $6.5 \times 10^{-3}$
        &
        $2.2 \times 10^{-2}$
        &
        $6.8 \times 10^{-2}$ \\ 
        \hline
        a: uniform (1 AU), e: uniform
        &
        $3.2 \times 10^{-4}$
        &
        $1.5 \times 10^{-2}$
        &
        $4.0 \times 10^{-2}$
        &
        0.12 \\ 
        \hline
        a: uniform (1 AU), e: thermal
        &
        $4.4 \times 10^{-4}$
        &
        $2.2 \times 10^{-2}$
        &
        $6.3 \times 10^{-2}$
        &
        0.15 \\ 
        \hline
        \textbf{SSE mass loss (0.001)} \\
        \hline
        a: log-uniform (0.1 AU), e: uniform
        &
        $2 \times 10^{-5}$$^*$
        &
        $6.1 \times 10^{-4}$
        &
        $2.5 \times 10^{-3}$
        &
        $1.6 \times 10^{-2}$ \\ 
        \hline
        a: log-uniform (0.1 AU), e: thermal
        &
        0$^*$
        &
        $1.2 \times 10^{-3}$
        &
        $5.0 \times 10^{-3}$
        &
        $3.0 \times 10^{-2}$ \\ 
        \hline
        a: uniform (0.1 AU), e: uniform
        &
        $2 \times 10^{-5}$$^*$
        &
        $8.8 \times 10^{-4}$
        &
        $3.6 \times 10^{-3}$
        &
        $2.3 \times 10^{-2}$ \\ 
        \hline
        a: uniform (0.1 AU), e: thermal
        &
        $4 \times 10^{-5}$$^*$
        &
        $1.8 \times 10^{-3}$
        &
        $6.4 \times 10^{-3}$
        &
        $4.2 \times 10^{-2}$ \\ 
        \hline
         a: log-uniform (1 AU), e: uniform
        &
        $4 \times 10^{-5}$
        &
        $6.4 \times 10^{-4}$
        &
        $2.5 \times 10^{-3}$
        &
        $1.6 \times 10^{-2}$ \\ 
        \hline
        a: log-uniform (1 AU), e: thermal
        &
        $2 \times 10^{-5}$
        &
        $1.1 \times 10^{-3}$
        &
        $4.5 \times 10^{-3}$
        &
        $3.0 \times 10^{-2}$ \\ 
        \hline
        a: uniform (1 AU), e: uniform
        &
        $2.9 \times 10^{-4}$
        &
        $1.4 \times 10^{-2}$
        &
        $4.1 \times 10^{-2}$
        &
        0.13 \\ 
        \hline
        a: uniform (1 AU), e: thermal
        &
        $7.0 \times 10^{-4}$
        &
        $2.2 \times 10^{-2}$
        &
        $6.4 \times 10^{-2}$
        &
        0.16 \\ 
        \hline
        \textbf{SSE mass loss (0.01)} \\
        \hline
        a: log-uniform (0.1 AU), e: uniform
        &
        0$^*$
        &
        $6.7 \times 10^{-4}$
        &
        $2.2 \times 10^{-3}$
        &
        $1.5 \times 10^{-2}$ \\ 
        \hline
        a: log-uniform (0.1 AU), e: thermal
        &
        $10^{-5}$$^*$
        &
        $1.2 \times 10^{-3}$
        &
        $4.5 \times 10^{-3}$
        &
        $2.8 \times 10^{-2}$ \\ 
        \hline
        a: uniform (0.1 AU), e: uniform
        &
        $2 \times 10^{-5}$$^*$
        &
        $8.5 \times 10^{-4}$
        &
        $3.1 \times 10^{-3}$
        &
        $2.1 \times 10^{-2}$ \\ 
        \hline
        a: uniform (0.1 AU), e: thermal
        &
        $2 \times 10^{-5}$$^*$
        &
        $1.7 \times 10^{-3}$
        &
        $6.0 \times 10^{-3}$
        &
        $3.8 \times 10^{-2}$ \\ 
        \hline
         a: log-uniform (1 AU), e: uniform
        &
        $10^{-5}$
        &
        $5.1 \times 10^{-4}$
        &
        $2.1 \times 10^{-3}$
        &
        $1.5 \times 10^{-2}$ \\ 
        \hline
        a: log-uniform (1 AU), e: thermal
        &
        $10^{-5}$
        &
        $1.2 \times 10^{-3}$
        &
        $4.2 \times 10^{-3}$
        &
        $2.8 \times 10^{-2}$ \\ 
        \hline
        a: uniform (1 AU), e: uniform
        &
        $2.5 \times 10^{-4}$
        &
        $1.3 \times 10^{-2}$
        &
        $4.0 \times 10^{-2}$
        &
        0.12 \\ 
        \hline
        a: uniform (1 AU), e: thermal
        &
        $4.4 \times 10^{-4}$
        &
        $2.2 \times 10^{-2}$
        &
        $6.4 \times 10^{-2}$
        &
        0.16 \\         
        \hline
    \end{tabular}
    \caption{Probability of producing systems with tilt greater than $\pi/2$ after the secondary SN kicks. Note that the (0.1 AU) or (1 AU) in each model is the upper bound for the semimajor axis range. Values with $*$ mark the models that are more physically motivated.}
    \label{table}
\end{table*}

In our analysis, we model the formation of GW191109 through isolated binary evolution using the following simple assumptions: 
\begin{itemize}
    \item The more massive star forms the more massive BH in the binary first, with no contribution to the final tilt of the orbital plane, as we assume that alignment of the BH spins and the binary orbital angular momentum occurred before the formation of the second BH.
    \item The kick from the second SN is the primary contribution to the tilt of the orbital plane with respect to the BBH spin axes.
    \item The natal kick imparted to the secondary BH at birth is isotropic.
\end{itemize}
With our methodology, we are able to conduct a controlled analysis of how different initial parameters affect the final distribution of tilts, thus avoiding all the uncertainties associated with modeling the full formation process. Moreover, the component masses of GW191109 (in particular the primary) lie in the mass gap, which can be formed in isolated binaries only when considering the uncertain combined effect of the hydrogen-rich envelopes, dredge-ups, and $^{12}{\rm C}(\alpha, \gamma)^{16}{\rm O}$ nuclear rates in massive stars \citep[e.g.,][]{farmer20, renzo20b, costa21}. Concerning spins, we expect that phases of mass transfer cause the alignment of the spin axes of the stars and BHs to the orbital angular momentum. As a result, the main contribution to the spin misalignment with respect to the orbital angular momentum (to eventually get a negative effective spin) is the natal kick imparted by an asymmetric SN explosion when forming the secondary BH.

To quantitatively assess the distributions of spin-orbit misalignment produced as a result of natal kicks, we compute the tilt of the binary orbit by comparing pre-SN and post-SN energy and angular momentum \citep{hills83, brandt95, kalogera96, kalogera00, pijloo12, fragione21}. We consider a binary of masses $m_{\rm BH,1}$ and $m_2$, semi-major axis $a$, and eccentricity $e$. The new orbital semimajor axis $a_{\rm n}$, eccentricity $e_{\rm n}$, and spin-orbit misalignment change as a result of a SN explosion of the secondary due to mass loss ($\Delta m_2=m_2-m_{\rm BH,2}$), and the isotropic natal kick $\bf{v_k}$ on the newly born BH is imparted. Assuming that the SN takes place instantaneously, at a relative separation $\bf{r}$ and velocity $\bf{v}$, the misalignment between the post-SN and pre-SN orbit is  
\begin{equation}
\Delta \theta=\arccos \left(\frac{\bf{h}\cdot \bf{h_{\rm n}}}{h\ h_{\rm n}} \right) \,,
\end{equation}
where
\begin{equation}
|{\bf{h}}|^2=|{\bf{r}}\times {\bf{v}}|^2=G(m_{\rm BH,1}+m_2) a(1-e)^2\,,
\label{eqn:hcons1}
\end{equation}
is the pre-SN angular momentum, and
\begin{equation}
|{\bf{h_n}}|^2=|{\bf{r}}\times {\bf{v_{\rm n}}}|^2=G(m_{\rm BH,1}+m_{\rm BH,2}) a_{\rm n}(1-e_{\rm n})^2\,,
\label{eqn:hcons2}
\end{equation}
is the post-SN angular momentum, with ${\bf{v_n}}={\bf{v}}+{\bf{v_k}}$ being the new relative velocity.

Since we are modeling binary systems that result in a GW191109-like merger, we set the primary mass to be $65 M_{\odot}$ and the post-SN secondary mass to be $47 M_{\odot}$, taking the medians of their respective parameter estimation distributions. For the remaining parameters, we test a range of initial conditions for a total of 96 different models:

\begin{itemize}
    \item Initial semimajor axis: We test both log-uniform and uniform distributions with ranges $10^{-2}$~AU - $10^{-1}$~AU and $10^{-2}$~AU - $1$~AU, for a total of 4 distinct semimajor axis distributions. 
    \item Initial eccentricity: We take into account both uniform and thermal distributions, for a total of 2 distinct eccentricity distributions.
    \item Pre-SN mass of $m_2$: We consider 3 distinct pre-SN mass distributions for $m_2$. The first pre-SN mass distribution we test is a uniform distribution in the range $47.5 \rm M_{\odot}$ - $57 \rm M_{\odot}$. The other two pre-SN mass distributions are simulated distributions generated using the Single Star Evolution (SSE) code with metallicities 0.001 and 0.01 \citep{hurley00}. For the SSE simulations, we run a population of single stars to Hubble time and extract only stars with post-SN masses within the mass error range of the GW191109 secondary mass: $(34\rm M_{\odot}, 62\rm M_{\odot})$. Since it is unfeasible to obtain stars with an exact mass of $47 \rm M_{\odot}$ through SSE, we consider LVK's mass error range and assume that the mass loss experienced by these stars is representative of a $47 \rm M_{\odot}$ star. We show the SSE-simulated mass loss distributions in Appendix \ref{fig:a1} and \ref{fig:a2}. We add these mass--loss values to the GW191109 secondary mass median value, $47\rm M_{\odot}$, to get the two SSE pre-SN mass distributions. 
    \item Kick velocity: We use a Maxwellian distribution
    \begin{equation}
        p(v_\mathrm{kick})\propto v_\mathrm{kick}^2 \exp \left ( -\frac{v_\mathrm{kick}^2}{2\sigma^2} \right ),
        \label{eqn:vkick}
    \end{equation}
    with velocity dispersion $\sigma$ to model the velocity kick imparted to the secondary BH at birth. Since $\sigma$ is highly uncertain, we consider 4 different models. We use $\sigma$ = 50 km s$^{-1}$, 100 km s$^{-1}$, and 265 km s$^{-1}$, for three of our models. Note that $\sigma=265$ km s$^{-1}$ is the typical kick expected for NSs as found by \citet{hobbs05}, while $\sigma=100$ km s$^{-1}$ as in the analysis of \citet{arzoumanian02}. For the fourth model, we adopt natal kicks from Eq.~\ref{eqn:vkick} as for NSs, but with $\sigma$ scaled by linear momentum conservation as follows \citep{repetto12, janka13}:
    \begin{equation}
        v_{\rm BH} = \frac{\langle m_{\rm NS} \rangle}{m_{\rm BH}} v_{\rm NS},
        \label{eqn:momconserv}
    \end{equation}
    where $v_{\rm BH}$ is the natal kick on a BH with mass $m_{\rm BH}$, $\langle m_{\rm NS} \rangle \approx 1.3 M_{\odot}$ is the average mass of NSs, and $v_{\rm NS}$ is randomly drawn from the Maxwellian distribution with $\sigma$ = 265 km s$^{-1}$. Here, we fix $m_{\rm BH}$ to $47\rm M_{\odot}$, the median mass of the secondary BH in GW191109. We label this model in Table \ref{table} as ``Momentum Conservation.'' We note that more recently, non-Maxwellian distributions have been used in the literature \citep{kapil23}, but we do not test these distributions in our modeling.
\end{itemize}
For each combination of parameter distributions, we simulate $10^5$ binaries for a total of 9.6 million binaries.

Our results primarily focus on the final tilt after the second SN. In order to have a negative $\chi_{\rm eff}$, the natal kick imparted to the secondary has to be strong enough to tilt the orbital angular momentum more than $\pi/2$ with respect to its initial orientation. In Figure~\ref{fig:2}, we show the cumulative distribution of the final tilts for binaries, with an initial semimajor axis distribution following a log-uniform distribution in range $10^{-2}$ AU - $10^{-1}$ AU, an initial uniform eccentricity distribution, and a SSE-simulated pre-SN mass distribution with metallicity $0.01$. The models in Figure \ref{fig:2} clearly demonstrate that it is unlikely to tilt binaries above $\pi/2$, and a direct correlation exists between higher velocity kicks and the fraction of binaries experiencing greater tilts. Specifically, no binaries experienced tilts greater than $\pi/2$ when the velocity kick distribution followed momentum conservation, and $6.7 \times 10^{-4}$, $2.16 \times 10^{-3}$, $1.535 \times 10^{-2}$ of the $10^5$ binaries experienced tilts greater than $\pi/2$ for Maxwellian velocity kick distributions with $\sigma$ = 50 km s$^{-1}$, $\sigma$ = 100 km s$^{-1}$, and $\sigma$ = 265 km s$^{-1}$, respectively. 

\begin{figure*}
    \centering
    \includegraphics[width=\textwidth]{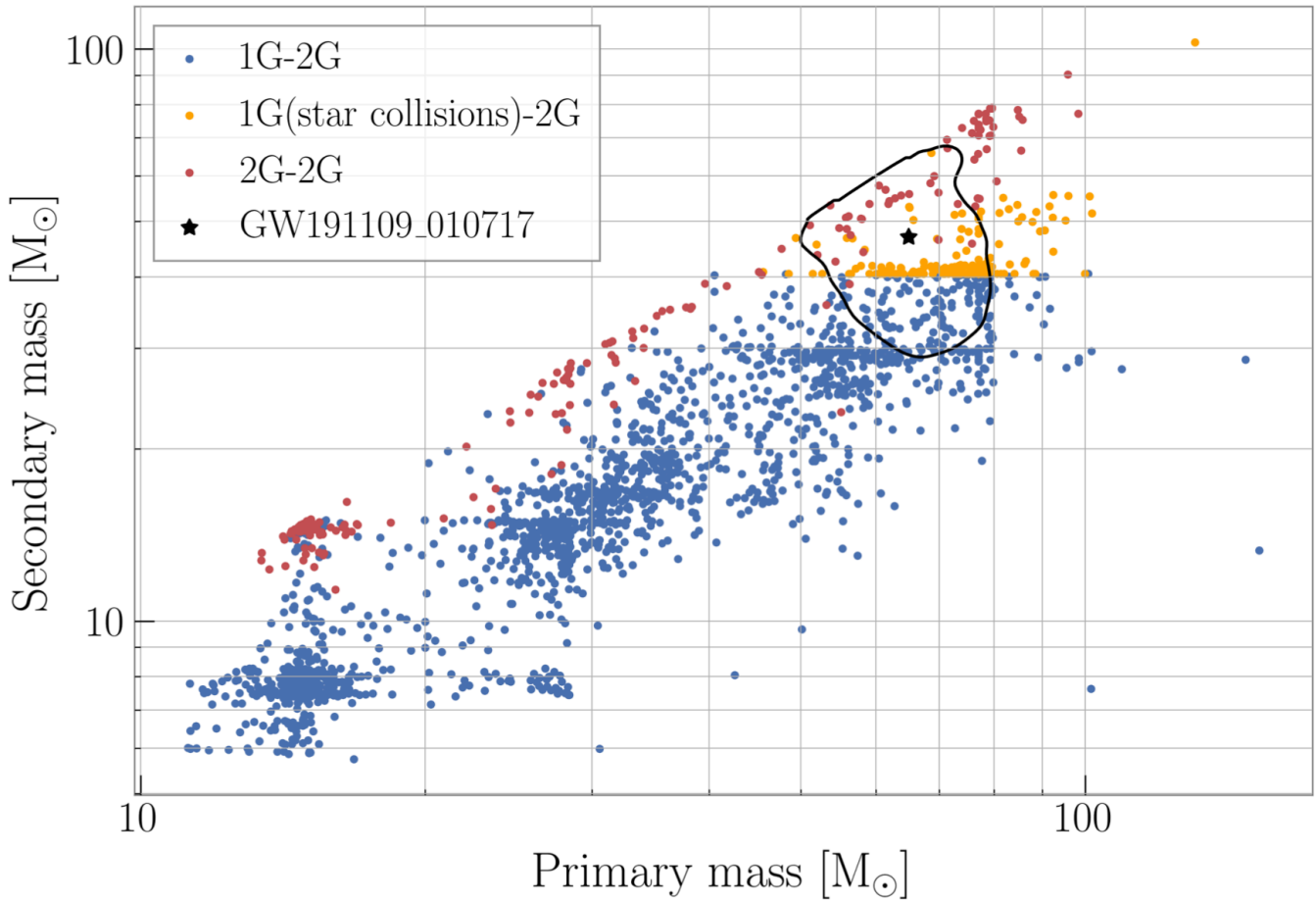}
    \caption{Primary mass and secondary mass of 1G-2G and 2G-2G BBHs from 148 simulated globular clusters in the models of \citet{kremer20}. For 1G-2G mergers, we distinguish between 1G BHs born as a result of the collapse of massive stars (blue points) and 1G BHs formed via star collisions (yellow points). We label GW191109's median component masses with a black star, with the contour line corresponding to its $90\%$ confidence region in mass space.}
    \label{fig:fig3}
\end{figure*}

The results for all 96 models are summarized in  Table \ref{table}. We place a star next to the values corresponding to models that are more likely to be representative of a population of merging BBHs. One of the factors that we consider is whether or not the BBHs can merge in Hubble time to be observed today, given the merger timescale \citep{peters64}
\begin{eqnarray}
    T_{\rm GW} & \approx & 13  {\rm Gyr} \left( \frac{2 \times 10^3 \rm \msun^3}{m_{\rm BH,1}m_{\rm BH,2}(m_{\rm BH,1}+m_{\rm BH,2})} \right) \nonumber\\
    & \times & \left( \frac{a_n}{0.095 \rm AU} \right)^4 (1-e_n^2)^{7/2}\,.
\end{eqnarray}
This equation places a limit on how large the semimajor axis can be in order for a BBH to merge in Hubble time. Specifically, given $m_{\rm BH,1}$ and $m_{\rm BH,1}$ as 65 and 47 $M_\odot$, respectively, a semimajor axis of 1 AU would require at least a 0.84 eccentricity for the merger timescale to be less than a Hubble time. On the other hand, a semimajor axis of 0.1 AU would allow for the full range of eccentricities. Thus, a uniform or log-uniform semimajor distribution ranging from $0.01$ to $0.1$ AU is more likely to be representative of a population of merging BBHs than that of the range $0.01$ to $1$ AU. For completeness, we show all of these models in Table \ref{table}, but we star the values with an upper bound of 0.1 AU for the semimajor axis distribution. Another factor to consider what more likely represents a population of merging BBHs is the velocity kick distribution. A Maxwellian distribution with a $\sigma$ value of 265 km s$^{-1}$ is the expected velocity kick distribution for neutron stars, but is very unlikely to be the case for much more massive BHs, such as in this case due to momentum conservation \citep{belczynski08}. Thus, we star the numbers in Table \ref{table} that assume a momentum conservation velocity kick distribution. Looking at the starred values in Table \ref{table}, we find that at most, only $0.004\%$ of isolated binaries can be tilted enough such that negative effective inspiral spin is produced.  

Due to the improbable chance of creating a binary with both a negative $\chi_{\rm eff}$ and BH component masses in the mass gap \citep[e.g.,][]{belczynski20,farmer20,renzo20b,costa21}, the rates for this formation channel would therefore be too low to explain GW191109. Thus, we conclude that isolated binary evolution is very unlikely to produce GW191109-like events. 

\subsection{Dynamics in dense star clusters}

Merging BBHs can be assembled dynamically in dense star clusters. Here, BHs quickly segregate to the center through dynamical friction \citep{chandrasekhar43, spitzer87} and interact with each other to form binaries. If the binary is hard, subsequent three- and four-body encounters with other BHs further harden the binary by extracting orbital energy \citep{heggie75}, which can eventually merge \citep[e.g.,][]{mcmillan91, hut92,portegies00,rodriguez16b,fragione18b,samsing18d,banerjee20}. 

An interesting possibility for BBH mergers formed through dynamical interactions is that the merger remnant can be retained in the parent cluster, whenever the relativistic recoil kick imparted as a result of asymmetric emission of GWs is smaller than the cluster escape speed \citep[e.g.,][]{lousto10,lousto12}. The retained BH, which is now a second-generation BH (2G), can be again dynamically processed to form a new BBH system that eventually merges. Higher-generation mergers account for $\sim 10\%$ of mergers from globular clusters, with this fraction increasing for denser environments such as nuclear star clusters \citep[e.g.,][]{rodriguez19,antonini19,mapelli21,fragione22, fragione22b}. From the second LIGO-Virgo GW Transient Catalog, \cite{kimball21} finds that the catalog contains at least one second-generation merger with 99\% credibility and lists five BBH mergers with high odds of involving at least one second-generation BH. 

In Figure~\ref{fig:fig3}, we show the component masses of the merging BBH extracted from the public globular cluster models of \citet{kremer20}. Out of $148$ globular cluster models, we find that there are 169 1G-2G and 24 2G-2G BBHs that have primary and secondary masses consistent with the 90\% confidence interval mass ranges of GW191109 \citep{ligo21}. Note that some of the 1G BHs have masses well within the mass gap since they originate from the collapse of a very massive stars, born as a result of repeated stellar mergers \citep{gonzalez21}. Therefore, 1G-2G and 2G-2G BBHs dynamically formed in dense star clusters can easily produce the masses of GW191109-like events, despite lying in the pair instability mass gap, unlike isolated binaries.

\begin{figure}
    \centering
    \includegraphics[width=0.475\textwidth]{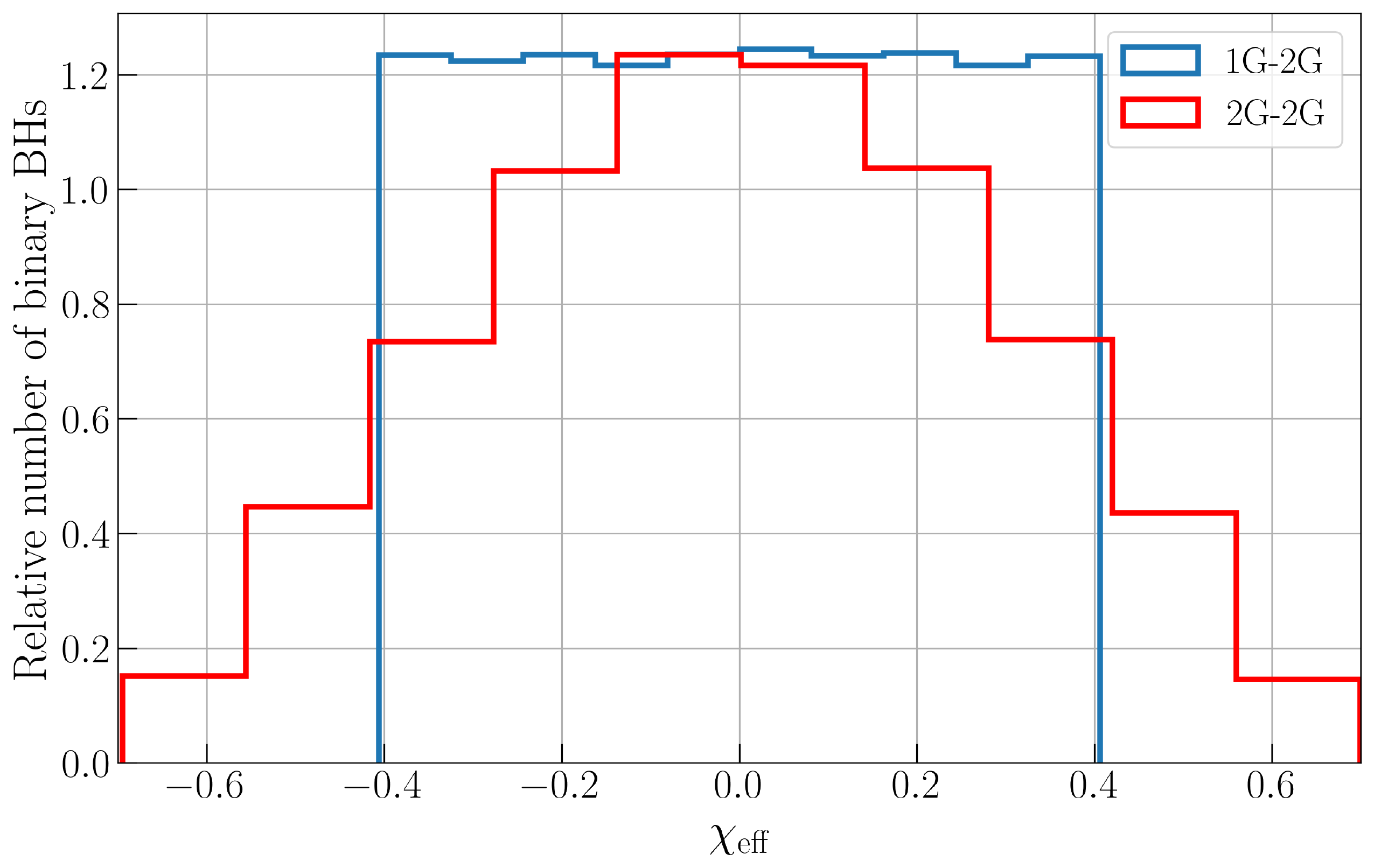}
    \caption{Distribution of effective spin for 1G-2G and 2G-2G BBHs, for component masses consistent with GW191109 and assuming non-spinning 1G BHs. Both distributions are symmetric around zero, with a non-negligible likelihood to reproduce negative $\chi_{\rm eff}$ values.}
    \label{fig:fig4}
\end{figure}

For dynamically assembled BBHs, the population of BH spins relative to the orbital angular momentum axis is isotropic, as a result of the few-body encounters that catalyze the merger of BBHs. We assume that 1G BHs are born with negligible spin \citep{qin18, fuller19} and 2G BHs have spins of about $0.7$ \citep[e.g.,][]{buonanno08, tichy08}. Combining the isotropically distributed spin directions with the component masses of GW191109, we get the $\chi_{\rm eff}$ distributions for both 1G-2G BBHs and 2G-2G BBHs, as shown in Figure~\ref{fig:fig4}. Since the distributions are symmetric around zero, with the 1G-2G distribution uniformly distributed between -0.41 and 0.41 and the 2G-2G distribution isotropically distributed about 0 between -0.70 and 0.70, negative values of $\chi_{\rm eff}$ are as likely as positive values. Under the assumption that 1G BHs are born with negligible spin, the median $\chi_{\rm eff}$ value of GW191109 of $-0.29$ can be easily accounted for with dynamically assembled BBHs as long as one of the components is a 2G BH. 

\begin{figure}
    \centering
    \includegraphics[width=0.475\textwidth]{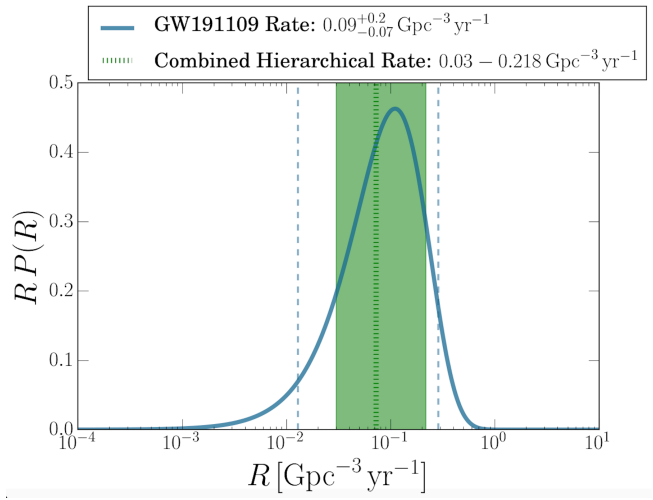}
    \caption{In solid blue, we plot the posterior over the rate of GW191109-like mergers assuming one Poisson-distributed count from a population drawn from the public parameter estimation samples. The dashed blue lines indicate the 90\% confidence interval. The dotted green line marks the combined 1G-2G and 2G-2G merger rate, assuming $\rm \rho_{GC}$ = 0.77 Mpc $^{-3}$, and the shaded green region marks the range of rates bounded by assuming  $\rm \rho_{GC}$ = 0.32 Mpc $^{-3}$ and  $\rm \rho_{GC}$ = 2.31 Mpc $^{-3}$.}
    \label{fig:fig5}
\end{figure}

Finally, we use the number of merging BBH consistent with GW191109 to compute the rates of GW191109-like events
\begin{equation}
    R = \frac{N \rho_{\rm GC}}{\rm \tau_H},
\end{equation}
where $N$ is the number of BBHs in the GW191109 mass range per globular cluster, $\rm \rho_{GC}$ is the density of globular clusters in the Universe, and $\rm \tau_H$ is Hubble time. We extract N from Figure~\ref{fig:fig3} by taking the number of simulated 1G-2G and 2G-2G events with masses within the GW191109 90\% confidence interval range and dividing it by the total number of simulated globular clusters, which is 169/148 and 24/148, respectively. We assume a value of 0.77 Mpc$^{-3}$ for $\rm \rho_{GC}$ with an optimistic value of 2.31 Mpc$^{-3}$ and pessimistic value of 0.32 Mpc$^{-3}$, using the calculations from \cite{rodriguez15}. In Figure \ref{fig:fig5}, we indicate the rate calculation assuming $\rm \rho_{GC}$ = 0.77 Mpc $^{-3}$ with the vertical green dotted line, and the shaded region accounts for the range of calculated rates bounded by the optimistic and pessimistic values for $\rm \rho_{GC}$. Assuming $\rm \rho_{GC}$ = 0.77 Mpc$^{-3}$, we get an approximate rate for 1G-2G events of 0.064 Gpc$^{-3}$ yr$^{-1}$ and for 2G-2G events of 0.0090 Gpc$^{-3}$ yr$^{-1}$, leading to a total rate of 0.073 Gpc$^{-3}$ yr$^{-1}$ for 1G-2G or 2G-2G events having masses in the GW191109 90\% confidence interval range. For an optimistic value of $\rm \rho_{GC}$ = 2.31 Mpc$^{-3}$, that total rate rises to 0.218 Gpc$^{-3}$ yr$^{-1}$, and for a pessimistic value of $\rm \rho_{GC}$ = 0.32 Mpc$^{-3}$, that total rate decreases to 0.030 Gpc$^{-3}$ yr$^{-1}$.

We then compare our estimate to the expected number of GW191109-like mergers given one detection by the LIGO-Virgo detector network. Following the method described in \citet{kim03}, we calculate this rate assuming one Poisson-distributed count from a BBH population with masses and spins drawn from the publicly available parameter estimation samples for GW191109 \citep{ligo21}. We calculate the network sensitivity to this population across O1, O2, and O3 using a combined SNR of 10 as a detection threshold. Using an $R^{-1/2}$ Jeffrey's prior, we find a GW191109-like merger rate of $R_{191109}=0.09_{-0.07}^{+0.2}$ Gpc$^{-3}$ yr$^{-1}$ (90\% confidence interval), and plot the rate posterior in Figure \ref{fig:fig5}. Here, we see that the simulated combined hierarchical rate lies well within the rate distribution of detecting a GW191109-like merger.

In summary, BBHs assembled in dense star clusters can explain the masses, the effective spin, and the rate of GW191109, unlike isolated binary evolution. Note also that our interpretation of GW191109 as having a dynamical origin is consistent with the findings in \cite{romero-shaw22}. In this study, they used a reweighting method \citep{payne19, romero-shaw19} to calculate the  eccentricity posterior probability distribution and find $72.19\%$ of its posterior support an eccentricity above 0.05, and $62.63\%$ above 0.1. This likely non-zero orbital eccentricity also points towards a dynamical origin of GW191109, consistent with our results. 

\subsection{Other channels}

Another astrophysical scenario to consider for the possible formation of GW191109-like events is the AGN channel \citep[e.g.,][]{bartos17,tagawa20}. The spin evolution of stellar-mass BHs in AGN disks has been studied by \cite{tagawa20c}. Through their semi-analytical simulations, they find that while gas accretion enhances the effective spin towards more positive values, hard binary-single interactions in the disks may reduce it. Therefore, the more efficient radial migration of BHs to inner, densely populated regions of the AGN disk is, the more symmetric around zero is the distribution of effective spin of merging BBHs. However, if this migration is inefficient, then the $\chi_{\rm eff}$ values would be skewed towards higher values. \cite{tagawa20} finds that BBH mergers with component masses in the mass-gap can be reproduced in the AGN channel, but their rates are highly uncertain. Considering the effective inspiral spin, masses, and rates, we cannot confidently say that GW191109 is likely to be formed via the AGN disk channel.

Hierarchical triple systems have been extensively studied in the literature \citep[e.g.,][]{antonini12,hoang18,hamers21}. Using population synthesis of triple stars that form a BH triplet, \citet{martinez22} found that the distribution of effective spins for merging BBH is likely symmetric around zero, as a result of the Lidov-Kozai oscillations. Therefore, hierarchical systems can produce mergers with a negative value of $\chi_{\rm eff}$. However, this channel is unlikely to produce mass-gap BBHs for the same reasons isolated binaries cannot, that is the masses of the component BHs are limited by the pair-instability physics. Finally, \cite{martinez22} finds that although highly uncertain, the hierarchical triple rates are estimated to possibly account for only a fraction of the observed BBH rates, which renders this scenario unlikely for GW191109-like events. 

Because these other channels are unlikely to be the origin of GW191109-like events, we argue that dynamical assembly in dense star clusters is the most likely formation channel for GW191109.

\section{Conclusions and Discussion}
\label{sec:sec4}

In this paper, we have demonstrated that GW191109 is unlikely to originate from isolated binary evolution based on its negative effective inspiral spin, masses, and inferred rate. Other channels such as hierarchical systems and AGN disks are also unlikely to explain GW191109. Combined with the possibility of non-zero eccentricity \citep{romero-shaw22}, all the evidence points towards a dynamical origin for this source.

Furthermore, this result can be extended to binary systems of any mass with some uncertainties. \cite{kalogera00} applied a similar prescription for BBH binaries, testing much lower masses in the 5~$\rm M_{\odot}$-20~$\rm M_{\odot}$ range and found that even by assuming a Maxwellian distribution with $\sigma$ = 200 km s$^{-1}$ for the velocity kicks, it is extremely unlikely for isolated binary evolution to tilt a BH over $\pi/2$ to contribute to a negative effective inspiral spin. We have reproduced similar results using our prescriptions for the different models with the BBH masses attained from GW200225, 19.3~$\rm M_{\odot}$ and 14~$\rm M_{\odot}$. This GW event, as shown in Figure \ref{fig:fig1}, also has a very likely negative effective inspiral spin. For the models involving a semimajor axis distribution with an upper bound of 0.1 AU and velocity kicks following momentum conservation, no more than $0.09\%$ of all binaries can be tilted over $\pi/2$ to produce a negative effective inspiral spin, demonstrating that even regardless of mass, it is extremely unlikely for isolated binary evolution to produce a negative effective inspiral spin. We note that a recent alternative theory for isolated binary evolution producing negative effective inspiral spins was suggested by \cite{tauris22}, as a result of the spin-axis tossing of BHs during their formation process in the core collapse of a massive star.

To study isolated binary evolution in our simulations, we have modeled how a binary system is affected after the second SN occurs in the system. It is possible that the first SN could also help tilt the binary system, increasing the total tilt of the BBHs' spins with respect to the angular momentum vector. While this is true, this contribution to the tilt will likely be smaller, as the binary was more massive during the first SN, making it more difficult to tilt the orbital plane. Given that the second SN is already unlikely to tilt the binary systems over $\pi/2$, the kick the BH experiences by the first SN would be even more unlikely to contribute to such a tilt. 

Finally, it is worth noting that in LIGO/Virgo's O3 run, approximately $20\%$ of the signals experienced glitches. The official LVK release data took into account this effect by modeling the glitch in their inference estimation. In the case of GW191109, the glitch was experienced by the Livingston data between 20 and 40 Hz. If these data are entirely removed from the inference, \cite{davis22} finds that the $\chi_{\rm eff}$ distribution does not peak strongly in the negative values anymore. While this information does not imply that a conclusion about the true $\chi_{\rm eff}$ distribution can be drawn, it does emphasize the uncertainties for the measurement. Additionally, we note that the LVK data analysis pipeline assumes a prior of isotropic spins for all of the GW events \citep{ligo21}, which may affect the posterior $\chi_{\rm eff}$ distribution. Even if the information on $\chi_{\rm eff}$ of GW191109 is excluded from the analysis of its origin, its masses and rates still strongly point towards a dynamical origin. 

With the LVK O4 run coming up in the next year, the population of BBHs will increase to hundreds of events, with the potential of providing better constraints on possible formation channels. The observable properties of masses in the pair instability mass gap, unequal mass ratios, and negative $\chi_{\rm eff}$, and inferred properties of non-zero orbital eccentricity and merger rates, will be large indicators of alternative formation channels from standard isolated binary evolution, such as dynamical origins. Thus, it will be important to place an emphasis on the analysis of these properties in future observations.

\begin{acknowledgments}

We thank Fred Rasio, Salvatore Vitale, and Katerina Chatziioannou for useful comments on an earlier version of the manuscript. G.F.\ acknowledges support from NASA Grant 80NSSC21K1722. C.K. is grateful for support from the Riedel Family Fellowship. V.K. was partially supported through a CIFAR Senior Fellowship and from Northwestern University, including the Daniel I. Linzer Distinguished University Professorship fund. 

\end{acknowledgments}
%




\appendix
\restartappendixnumbering

\section{Relative mass loss distributions of SSE simulations}
\label{append}
\begin{figure}
    \centering
    \includegraphics[width=0.475\textwidth]{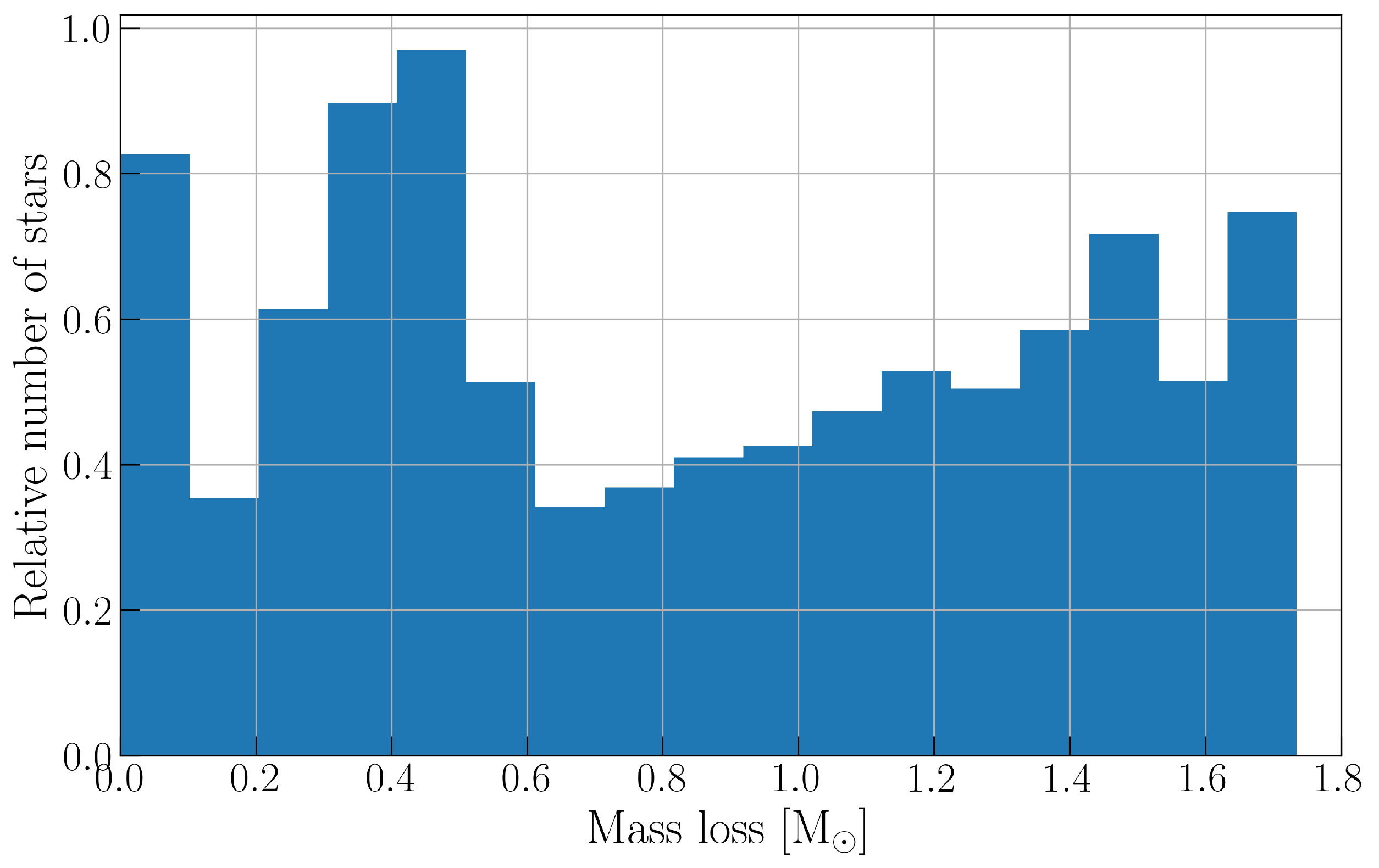}
    \caption{The relative mass loss distribution for a SSE-simulated population of single stars with initial metallicities of 0.001 and final masses within the error bar range of the GW191109 secondary mass.}
    \label{fig:a1}
\end{figure}

\begin{figure}
    \centering
    \includegraphics[width=0.475\textwidth]{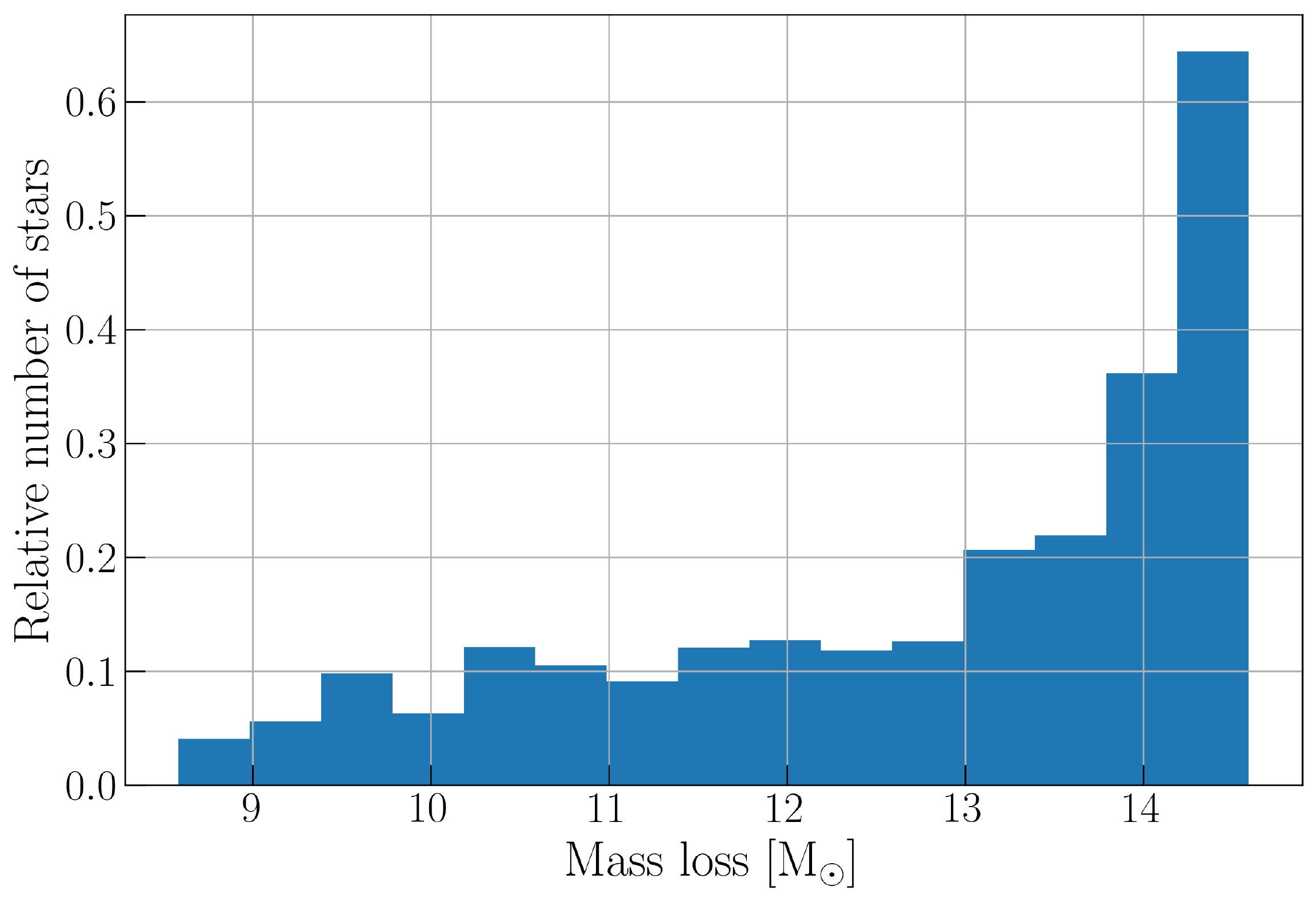}
    \caption{The relative mass loss distribution for a SSE-simulated population of single stars with initial metallicities of 0.01 and final masses within the error bar range of the GW191109 secondary mass.}
    \label{fig:a2}
\end{figure}

In this appendix, we show the relative mass loss distributions generated from SSE as explained in Section \ref{sec:sec3}. We simulate the evolution of $10^5$ single stars using SSE and plot the mass losses experienced from SN of those stars that have a post-SN mass that is within the error bar range of the GW191109 secondary mass. Figure \ref{fig:a1} and Figure \ref{fig:a2} show the relative distributions of mass loss given initial star metallicities of 0.001 and 0.01, respectively.  

\bibliography{sample631}{}

\begin{thebibliography}{}
\expandafter\ifx\csname natexlab\endcsname\relax\def\natexlab#1{#1}\fi
\providecommand{\url}[1]{\href{#1}{#1}}
\providecommand{\dodoi}[1]{doi:~\href{http://doi.org/#1}{\nolinkurl{#1}}}
\providecommand{\doeprint}[1]{\href{http://ascl.net/#1}{\nolinkurl{http://ascl.net/#1}}}
\providecommand{\doarXiv}[1]{\href{https://arxiv.org/abs/#1}{\nolinkurl{https://arxiv.org/abs/#1}}}

\bibitem[{{Abbott} {et~al.}(2020{\natexlab{a}})}]{lvc2020catb}
{Abbott}, R., {et~al.} 2020{\natexlab{a}}, arXiv e-prints, arXiv:2010.14533.
\newblock \doarXiv{2010.14533}

\bibitem[{{Abbott} {et~al.}(2020{\natexlab{b}})}]{lvc2020catc}
---. 2020{\natexlab{b}}, arXiv e-prints, arXiv:2010.14529.
\newblock \doarXiv{2010.14529}

\bibitem[{{Abbott} {et~al.}(2021){Abbott}, {Abbott}, {Abraham}, {Acernese},
  {Ackley}, {Adams}, {Adams}, {Adhikari}, {Adya}, {Affeldt}, {Agathos},
  {Agatsuma}, {Aggarwal}, {Aguiar}, {Aiello}, {Ain}, {Ajith}, {Akcay}, {Allen},
  {Allocca}, {Altin}, {Amato}, {Anand}, {Ananyeva}, {Anderson}, {Anderson},
  {Angelova}, {Ansoldi}, {Antelis}, {Antier}, {Appert}, {Arai}, {Araya},
  {Areeda}, {Ar{\`e}ne}, {Arnaud}, {Aronson}, {Arun}, {Asali}, {Ascenzi},
  {Ashton}, {Aston}, {Astone}, {Aubin}, {Aufmuth}, {AultONeal}, {Austin},
  {Avendano}, {Babak}, {Badaracco}, {Bader}, {Bae}, {Baer}, {Bagnasco},
  {Baird}, {Ball}, {Ballardin}, {Ballmer}, {Bals}, {Balsamo}, {Baltus},
  {Banagiri}, {Bankar}, {Bankar}, {Barayoga}, {Barbieri}, {Barish}, {Barker},
  {Barneo}, {Barnum}, {Barone}, {Barr}, {Barsotti}, {Barsuglia}, {Barta},
  {Bartlett}, {Bartos}, {Bassiri}, {Basti}, {Bawaj}, {Bayley}, {Bazzan},
  {Becher}, {B{\'e}csy}, {Bedakihale}, {Bejger}, {Belahcene}, {Beniwal},
  {Benjamin}, {Bennett}, {Bentley}, {Bergamin}, {Berger}, {Bergmann},
  {Bernuzzi}, {Berry}, {Bersanetti}, {Bertolini}, {Betzwieser}, {Bhandare},
  {Bhandari}, {Bhattacharjee}, {Bidler}, {Bilenko}, {Billingsley}, {Birney},
  {Birnholtz}, {Biscans}, {Bischi}, {Biscoveanu}, {Bisht}, {Bitossi},
  {Bizouard}, {Blackburn}, {Blackman}, {Blair}, {Blair}, {Blair}, {Blanch},
  {Bobba}, {Bode}, {Boer}, {Boetzel}, {Bogaert}, {Boldrini}, {Bondu},
  {Bonilla}, {Bonnand}, {Booker}, {Boom}, {Bork}, {Boschi}, {Bose},
  {Bossilkov}, {Boudart}, {Bouffanais}, {Bozzi}, {Bradaschia}, {Brady},
  {Bramley}, {Branchesi}, {Brau}, {Breschi}, {Briant}, {Briggs}, {Brighenti},
  {Brillet}, {Brinkmann}, {Brockill}, {Brooks}, {Brooks}, {Brown}, {Brunett},
  {Bruno}, {Bruntz}, {Buikema}, {Bulik}, {Bulten}, {Buonanno}, {Buscicchio},
  {Buskulic}, {Byer}, {Cabero}, {Cadonati}, {Caesar}, {Cagnoli}, {Cahillane},
  {Calder{\'o}n Bustillo}, {Callaghan}, {Callister}, {Calloni}, {Camp},
  {Canepa}, {Cannon}, {Cao}, {Cao}, {Carapella}, {Carbognani}, {Carney},
  {Carpinelli}, {Carullo}, {Carver}, {Casanueva Diaz}, {Casentini}, {Caudill},
  {Cavagli{\`a}}, {Cavalier}, {Cavalieri}, {Cella}, {Cerd{\'a}-Dur{\'a}n},
  {Cesarini}, {Chaibi}, {Chakravarti}, {Chan}, {Chan}, {Chandra}, {Chanial},
  {Chao}, {Charlton}, {Chase}, {Chassande-Mottin}, {Chatterjee},
  {Chattopadhyay}, {Chaturvedi}, {Chatziioannou}, {Chen}, {Chen}, {Chen},
  {Chen}, {Cheng}, {Cheong}, {Chia}, {Chiadini}, {Chierici}, {Chincarini},
  {Chiummo}, {Cho}, {Cho}, {Cho}, {Choate}, {Christensen}, {Chu}, {Chua},
  {Chung}, {Chung}, {Ciani}, {Ciecielag}, {Cie{\'s}lar}, {Cifaldi}, {Ciobanu},
  {Ciolfi}, {Cipriano}, {Cirone}, {Clara}, {Clark}, {Clark}, {Clarke},
  {Clearwater}, {Clesse}, {Cleva}, {Coccia}, {Cohadon}, {Cohen}, {Colleoni},
  {Collette}, {Collins}, {Colpi}, {Constancio}, {Conti}, {Cooper}, {Corban},
  {Corbitt}, {Cordero-Carri{\'o}n}, {Corezzi}, {Corley}, {Cornish}, {Corre},
  {Corsi}, {Cortese}, {Costa}, {Cotesta}, {Coughlin}, {Coughlin}, {Coulon},
  {Countryman}, {Cousins}, {Couvares}, {Covas}, {Coward}, {Cowart}, {Coyne},
  {Coyne}, {Creighton}, {Creighton}, {Croquette}, {Crowder}, {Cudell},
  {Cullen}, {Cumming}, {Cummings}, {Cunningham}, {Cuoco}, {Cury{\l}o},
  {Canton}, {D{\'a}lya}, {Dana}, {DaneshgaranBajastani}, {D'Angelo}, {Danila},
  {Danilishin}, {D'Antonio}, {Danzmann}, {Darsow-Fromm}, {Dasgupta}, {Datrier},
  {Dattilo}, {Dave}, {Davier}, {Davies}, {Davis}, {Daw}, {Dean}, {DeBra},
  {Deenadayalan}, {Degallaix}, {De Laurentis}, {Del{\'e}glise}, {Del Favero},
  {De Lillo}, {De Lillo}, {Del Pozzo}, {DeMarchi}, {De Matteis}, {D'Emilio},
  {Demos}, {Denker}, {Dent}, {Depasse}, {De Pietri}, {De Rosa}, {De Rossi},
  {DeSalvo}, {de Varona}, {Dhurandhar}, {D{\'\i}az}, {Diaz-Ortiz}, {Didio},
  {Dietrich}, {Di Fiore}, {DiFronzo}, {Di Giorgio}, {Di Giovanni}, {Di
  Giovanni}, {Di Girolamo}, {Di Lieto}, {Ding}, {Di Pace}, {Di Palma}, {Di
  Renzo}, {Divakarla}, {Dmitriev}, {Doctor}, {D'Onofrio}, {Donovan}, {Dooley},
  {Doravari}, {Dorrington}, {Downes}, {Drago}, {Driggers}, {Du}, {Ducoin},
  {Dupej}, {Durante}, {D'Urso}, {Duverne}, {Dwyer}, {Easter}, {Eddolls},
  {Edelman}, {Edo}, {Edy}, {Effler}, {Eichholz}, {Eikenberry}, {Eisenmann},
  {Eisenstein}, {Ejlli}, {Errico}, {Essick}, {Estell{\'e}s}, {Estevez},
  {Etienne}, {Etzel}, {Evans}, {Evans}, {Ewing}, {Fafone}, {Fair}, {Fairhurst},
  {Fan}, {Farah}, {Farinon}, {Farr}, {Farr}, {Fauchon-Jones}, {Favata}, {Fays},
  {Fazio}, {Feicht}, {Fejer}, {Feng}, {Fenyvesi}, {Ferguson},
  {Fernandez-Galiana}, {Ferrante}, {Ferreira}, {Fidecaro}, {Figura}, {Fiori},
  {Fiorucci}, {Fishbach}, {Fisher}, {Fishner}, {Fittipaldi}, {Fitz-Axen},
  {Fiumara}, {Flaminio}, {Floden}, {Flynn}, {Fong}, {Font}, {Forsyth},
  {Fournier}, {Frasca}, {Frasconi}, {Frei}, {Freise}, {Frey}, {Frey},
  {Fritschel}, {Frolov}, {Fronz{\'e}}, {Fulda}, {Fyffe}, {Gabbard}, {Gadre},
  {Gaebel}, {Gair}, {Gais}, {Galaudage}, {Gamba}, {Ganapathy}, {Ganguly},
  {Gaonkar}, {Garaventa}, {Garc{\'\i}a-Quir{\'o}s}, {Garufi}, {Gateley},
  {Gaudio}, {Gayathri}, {Gemme}, {Gennai}, {George}, {George}, {George},
  {Gergely}, {Ghonge}, {Ghosh}, {Ghosh}, {Ghosh}, {Giacomazzo}, {Giacoppo},
  {Giaime}, {Giardina}, {Gibson}, {Gier}, {Gill}, {Giri}, {Glanzer}, {Gleckl},
  {Godwin}, {Goetz}, {Goetz}, {Gohlke}, {Goncharov}, {Gonz{\'a}lez},
  {Gopakumar}, {Gossan}, {Gosselin}, {Gouaty}, {Grace}, {Grado}, {Granata},
  {Granata}, {Grant}, {Gras}, {Grassia}, {Gray}, {Gray}, {Greco}, {Green},
  {Green}, {Gretarsson}, {Griggs}, {Grignani}, {Grimaldi}, {Grimes}, {Grimm},
  {Grote}, {Grunewald}, {Gruning}, {Guerrero}, {Guidi}, {Guimaraes},
  {Guix{\'e}}, {Gulati}, {Guo}, {Gupta}, {Gupta}, {Gupta}, {Gustafson},
  {Gustafson}, {Guzman}, {Haegel}, {Halim}, {Hall}, {Hamilton}, {Hammond},
  {Haney}, {Hanke}, {Hanks}, {Hanna}, {Hannam}, {Hannuksela}, {Hannuksela},
  {Hansen}, {Hansen}, {Hanson}, {Harder}, {Hardwick}, {Haris}, {Harms},
  {Harry}, {Harry}, {Hartwig}, {Hasskew}, {Haster}, {Haughian}, {Hayes},
  {Healy}, {Heidmann}, {Heintze}, {Heinze}, {Heinzel}, {Heitmann}, {Hellman},
  {Hello}, {Helmling-Cornell}, {Hemming}, {Hendry}, {Heng}, {Hennes}, {Hennig},
  {Hennig}, {Hernandez Vivanco}, {Heurs}, {Hild}, {Hill}, {Hines}, {Hochheim},
  {Hofgard}, {Hofman}, {Hohmann}, {Holgado}, {Holland}, {Hollows}, {Holmes},
  {Holt}, {Holz}, {Hopkins}, {Horst}, {Hough}, {Howell}, {Hoy}, {Hoyland},
  {Huang}, {H{\"u}bner}, {Huddart}, {Huerta}, {Hughey}, {Hui}, {Husa},
  {Huttner}, {Hutzler}, {Huxford}, {Huynh-Dinh}, {Idzkowski}, {Iess},
  {Imperato}, {Inchauspe}, {Ingram}, {Intini}, {Isi}, {Iyer},
  {JaberianHamedan}, {Jacqmin}, {Jadhav}, {Jadhav}, {James}, {Jani},
  {Janssens}, {Janthalur}, {Jaranowski}, {Jariwala}, {Jaume}, {Jenkins},
  {Jeunon}, {Jiang}, {Johns}, {Johnson-McDaniel}, {Jones}, {Jones}, {Jones},
  {Jones}, {Jones}, {Jonker}, {Ju}, {Junker}, {Kalaghatgi}, {Kalogera},
  {Kamai}, {Kandhasamy}, {Kang}, {Kanner}, {Kapadia}, {Kapasi}, {Karathanasis},
  {Karki}, {Kashyap}, {Kasprzack}, {Kastaun}, {Katsanevas}, {Katsavounidis},
  {Katzman}, {Kawabe}, {K{\'e}f{\'e}lian}, {Keitel}, {Key}, {Khadka},
  {Khalili}, {Khan}, {Khan}, {Khazanov}, {Khetan}, {Khursheed}, {Kijbunchoo},
  {Kim}, {Kim}, {Kim}, {Kim}, {Kim}, {Kim}, {Kimball}, {King}, {Kinley-Hanlon},
  {Kirchhoff}, {Kissel}, {Kleybolte}, {Klimenko}, {Knowles}, {Knyazev}, {Koch},
  {Koehlenbeck}, {Koekoek}, {Koley}, {Kolstein}, {Komori}, {Kondrashov},
  {Kontos}, {Koper}, {Korobko}, {Korth}, {Kovalam}, {Kozak}, {Kr{\"a}mer},
  {Kringel}, {Krishnendu}, {Kr{\'o}lak}, {Kuehn}, {Kumar}, {Kumar}, {Kumar},
  {Kumar}, {Kuns}, {Kwang}, {Lackey}, {Laghi}, {Lalande}, {Lam}, {Lamberts},
  {Landry}, {Lane}, {Lang}, {Lange}, {Lantz}, {Lanza}, {La Rosa},
  {Lartaux-Vollard}, {Lasky}, {Laxen}, {Lazzarini}, {Lazzaro}, {Leaci},
  {Leavey}, {Lecoeuche}, {Lee}, {Lee}, {Lee}, {Lee}, {Lehmann}, {Leon},
  {Leroy}, {Letendre}, {Levin}, {Li}, {Li}, {Li}, {Li}, {Li}, {Linde},
  {Linker}, {Linley}, {Littenberg}, {Liu}, {Liu}, {Llorens-Monteagudo}, {Lo},
  {Lockwood}, {London}, {Longo}, {Lorenzini}, {Loriette}, {Lormand}, {Losurdo},
  {Lough}, {Lousto}, {Lovelace}, {L{\"u}ck}, {Lumaca}, {Lundgren}, {Ma},
  {Macas}, {MacInnis}, {Macleod}, {MacMillan}, {Macquet}, {Maga{\~n}a
  Hernandez}, {Maga{\~n}a-Sandoval}, {Magazz{\~A}{\textonesuperior}}, {Magee},
  {Majorana}, {Maksimovic}, {Maliakal}, {Malik}, {Man}, {Mandic}, {Mangano},
  {Mansell}, {Manske}, {Mantovani}, {Mapelli}, {Marchesoni}, {Marion},
  {M{\'a}rka}, {M{\'a}rka}, {Markakis}, {Markosyan}, {Markowitz}, {Maros},
  {Marquina}, {Marsat}, {Martelli}, {Martin}, {Martin}, {Martinez}, {Martinez},
  {Martynov}, {Masalehdan}, {Mason}, {Massera}, {Masserot}, {Massinger},
  {Masso-Reid}, {Mastrogiovanni}, {Matas}, {Mateu-Lucena}, {Matichard},
  {Matiushechkina}, {Mavalvala}, {Maynard}, {McCann}, {McCarthy}, {McClelland},
  {McCormick}, {McCuller}, {McGuire}, {McIsaac}, {McIver}, {McManus}, {McRae},
  {McWilliams}, {Meacher}, {Meadors}, {Mehmet}, {Mehta}, {Melatos}, {Melchor},
  {Mendell}, {Menendez-Vazquez}, {Mercer}, {Mereni}, {Merfeld}, {Merilh},
  {Merritt}, {Merzougui}, {Meshkov}, {Messenger}, {Messick}, {Metzdorff},
  {Meyers}, {Meylahn}, {Mhaske}, {Miani}, {Miao}, {Michaloliakos}, {Michel},
  {Middleton}, {Milano}, {Miller}, {Millhouse}, {Mills}, {Milotti},
  {Milovich-Goff}, {Minazzoli}, {Minenkov}, {Mir}, {Mishkin}, {Mishra},
  {Mistry}, {Mitra}, {Mitrofanov}, {Mitselmakher}, {Mittleman}, {Mo},
  {Mogushi}, {Mohapatra}, {Mohite}, {Molina}, {Molina-Ruiz}, {Mondin},
  {Montani}, {Moore}, {Moraru}, {Morawski}, {Moreno}, {Morisaki}, {Mours},
  {Mow-Lowry}, {Mozzon}, {Muciaccia}, {Mukherjee}, {Mukherjee}, {Mukherjee},
  {Mukherjee}, {Mukund}, {Mullavey}, {Munch}, {Mu{\~n}iz}, {Murray}, {Nadji},
  {Nagar}, {Nardecchia}, {Naticchioni}, {Nayak}, {Neil}, {Neilson}, {Nelemans},
  {Nelson}, {Nery}, {Neunzert}, {Nitz}, {Ng}, {Ng}, {Nguyen}, {Nguyen},
  {Nguyen}, {Nichols}, {Nissanke}, {Nocera}, {Noh}, {North}, {Nothard},
  {Nuttall}, {Oberling}, {O'Brien}, {O'Dell}, {Oganesyan}, {Ogin}, {Oh}, {Oh},
  {Ohme}, {Ohta}, {Okada}, {Olivetto}, {Oppermann}, {Oram}, {O'Reilly},
  {Ormiston}, {Ortega}, {O'Shaughnessy}, {Ossokine}, {Osthelder}, {Ottaway},
  {Overmier}, {Owen}, {Pace}, {Pagano}, {Page}, {Pagliaroli}, {Pai}, {Pai},
  {Palamos}, {Palashov}, {Palomba}, {Pan}, {Panda}, {Pang}, {Pankow},
  {Pannarale}, {Pant}, {Paoletti}, {Paoli}, {Paolone}, {Parker}, {Pascucci},
  {Pasqualetti}, {Passaquieti}, {Passuello}, {Patel}, {Patricelli}, {Payne},
  {Pechsiri}, {Pedraza}, {Pegoraro}, {Pele}, {Penn}, {Perego}, {Perez},
  {P{\'e}rigois}, {Perreca}, {Perri{\`e}s}, {Petermann}, {Petterson},
  {Pfeiffer}, {Pham}, {Phukon}, {Piccinni}, {Pichot}, {Piendibene},
  {Piergiovanni}, {Pierini}, {Pierro}, {Pillant}, {Pilo}, {Pinard}, {Pinto},
  {Piotrzkowski}, {Pirello}, {Pitkin}, {Placidi}, {Plastino}, {Pluchar},
  {Poggiani}, {Polini}, {Pong}, {Ponrathnam}, {Popolizio}, {Porter},
  {Poverman}, {Powell}, {Pracchia}, {Prajapati}, {Prasai}, {Prasanna},
  {Pratten}, {Prestegard}, {Principe}, {Prodi}, {Prokhorov}, {Prosposito},
  {Prudenzi}, {Puecher}, {Punturo}, {Puosi}, {Puppo}, {P{\"u}rrer}, {Qi},
  {Quetschke}, {Quinonez}, {Quitzow-James}, {Raab}, {Raaijmakers}, {Radkins},
  {Radulesco}, {Raffai}, {Rafferty}, {Rail}, {Raja}, {Rajan}, {Rajbhandari},
  {Rakhmanov}, {Ramirez}, {Ramirez}, {Ramos-Buades}, {Rana}, {Rao},
  {Rapagnani}, {Rapol}, {Ratto}, {Raymond}, {Razzano}, {Read}, {Regimbau},
  {Rei}, {Reid}, {Reitze}, {Rettegno}, {Ricci}, {Richardson}, {Richardson},
  {Richardson}, {Ricker}, {Riemenschneider}, {Riles}, {Rizzo}, {Robertson},
  {Robinet}, {Rocchi}, {Rocha}, {Rodriguez}, {Rodriguez-Soto}, {Rolland},
  {Rollins}, {Roma}, {Romanelli}, {Romano}, {Romel}, {Romero}, {Romero-Shaw},
  {Romie}, {Ronchini}, {Rose}, {Rose}, {Rose}, {Rosell}, {Rosi{\'n}ska},
  {Rosofsky}, {Ross}, {Rowan}, {Rowlinson}, {Roy}, {Roy}, {Ruggi}, {Ryan},
  {Sachdev}, {Sadecki}, {Sadiq}, {Sakellariadou}, {Salafia}, {Salconi},
  {Saleem}, {Samajdar}, {Sanchez}, {Sanchez}, {Sanchez}, {Sanchis-Gual},
  {Sanders}, {Sandles}, {Santiago}, {Santos}, {Saravanan}, {Sarin}, {Sassolas},
  {Sathyaprakash}, {Sauter}, {Savage}, {Savant}, {Sawant}, {Sayah}, {Schaetzl},
  {Schale}, {Scheel}, {Scheuer}, {Schindler-Tyka}, {Schmidt}, {Schnabel},
  {Schofield}, {Sch{\"o}nbeck}, {Schreiber}, {Schulte}, {Schutz}, {Schwarm},
  {Schwartz}, {Scott}, {Scott}, {Seglar-Arroyo}, {Seidel}, {Sellers},
  {Sengupta}, {Sennett}, {Sentenac}, {Sequino}, {Sergeev}, {Setyawati},
  {Shaffer}, {Shahriar}, {Sharifi}, {Sharma}, {Sharma}, {Shawhan}, {Shen},
  {Shikauchi}, {Shink}, {Shoemaker}, {Shoemaker}, {Shukla}, {ShyamSundar},
  {Sieniawska}, {Sigg}, {Singer}, {Singh}, {Singh}, {Singha}, {Singhal},
  {Sintes}, {Sipala}, {Skliris}, {Slagmolen}, {Slaven-Blair}, {Smetana},
  {Smith}, {Smith}, {Somala}, {Son}, {Soni}, {Soni}, {Sorazu}, {Sordini},
  {Sorrentino}, {Sorrentino}, {Soulard}, {Souradeep}, {Sowell}, {Spencer},
  {Spera}, {Srivastava}, {Srivastava}, {Staats}, {Stachie}, {Steer},
  {Steinhoff}, {Steinke}, {Steinlechner}, {Steinlechner}, {Steinmeyer},
  {Stevenson}, {Stolle-McAllister}, {Stops}, {Stover}, {Strain}, {Stratta},
  {Strunk}, {Sturani}, {Stuver}, {S{\"u}dbeck}, {Sudhagar}, {Sudhir}, {Suh},
  {Summerscales}, {Sun}, {Sun}, {Sunil}, {Sur}, {Suresh}, {Sutton}, {Swinkels},
  {Szczepa{\'n}czyk}, {Tacca}, {Tait}, {Talbot}, {Tanasijczuk}, {Tanner},
  {Tao}, {Tapia}, {Tapia San Martin}, {Tasson}, {Taylor}, {Tenorio},
  {Terkowski}, {Thirugnanasambandam}, {Thomas}, {Thomas}, {Thomas}, {Thompson},
  {Thondapu}, {Thorne}, {Thrane}, {Tiwari}, {Tiwari}, {Tiwari}, {Toland},
  {Tolley}, {Tonelli}, {Tornasi}, {Torres-Forn{\'e}}, {Torrie}, {e Melo},
  {T{\"o}yr{\"a}}, {Tran}, {Trapananti}, {Travasso}, {Traylor}, {Tringali},
  {Tripathee}, {Trovato}, {Trudeau}, {Tsai}, {Tsang}, {Tse}, {Tso}, {Tsukada},
  {Tsuna}, {Tsutsui}, {Turconi}, {Ubhi}, {Udall}, {Ueno}, {Ugolini},
  {Unnikrishnan}, {Urban}, {Usman}, {Utina}, {Vahlbruch}, {Vajente}, {Vajpeyi},
  {Valdes}, {Valentini}, {Valsan}, {van Bakel}, {van Beuzekom}, {van den
  Brand}, {Van Den Broeck}, {Vander-Hyde}, {van der Schaaf}, {van Heijningen},
  {Vardaro}, {Vargas}, {Varma}, {Vass}, {Vas{\'u}th}, {Vecchio}, {Vedovato},
  {Veitch}, {Veitch}, {Venkateswara}, {Venneberg}, {Venugopalan}, {Verkindt},
  {Verma}, {Veske}, {Vetrano}, {Vicer{\'e}}, {Viets}, {Vijaykumar},
  {Villa-Ortega}, {Vinet}, {Vitale}, {Vo}, {Vocca}, {Vorvick}, {Vyatchanin},
  {Wade}, {Wade}, {Wade}, {Walet}, {Walker}, {Wallace}, {Wallace}, {Walsh},
  {Wang}, {Wang}, {Wang}, {Wang}, {Ward}, {Warner}, {Was}, {Washington},
  {Watchi}, {Weaver}, {Wei}, {Weinert}, {Weinstein}, {Weiss}, {Wellmann},
  {Wen}, {We{\ss}els}, {Westhouse}, {Wette}, {Whelan}, {White}, {White},
  {Whiting}, {Whittle}, {Wilken}, {Williams}, {Williams}, {Williamson},
  {Willis}, {Willke}, {Wilson}, {Wimmer}, {Winkler}, {Wipf}, {Woan}, {Woehler},
  {Wofford}, {Wong}, {Wrangel}, {Wright}, {Wu}, {Wysocki}, {Xiao}, {Yamamoto},
  {Yang}, {Yang}, {Yang}, {Yap}, {Yeeles}, {Yoon}, {Yu}, {Yu}, {Yuen},
  {Zadro{\.Z}ny}, {Zanolin}, {Zelenova}, {Zendri}, {Zevin}, {Zhang}, {Zhang},
  {Zhang}, {Zhang}, {Zhao}, {Zhao}, {Zheng}, {Zhou}, {Zhou}, {Zhu},
  {Zimmerman}, {Zlochower}, {Zucker}, {Zweizig}, {LIGO Scientific
  Collaboration}, \& {Virgo Collaboration}}]{ligo21a}
{Abbott}, R., {Abbott}, T.~D., {Abraham}, S., {et~al.} 2021, Physical Review X,
  11, 021053, \dodoi{10.1103/PhysRevX.11.021053}

\bibitem[{{Antonini} {et~al.}(2019){Antonini}, {Gieles}, \&
  {Gualandris}}]{antonini19}
{Antonini}, F., {Gieles}, M., \& {Gualandris}, A. 2019, \mnras, 486, 5008,
  \dodoi{10.1093/mnras/stz1149}

\bibitem[{{Antonini} \& {Perets}(2012)}]{antonini12}
{Antonini}, F., \& {Perets}, H.~B. 2012, \apj, 757, 27,
  \dodoi{10.1088/0004-637X/757/1/27}

\bibitem[{{Antonini} \& {Rasio}(2016)}]{antonini16}
{Antonini}, F., \& {Rasio}, F.~A. 2016, \apj, 831, 187,
  \dodoi{10.3847/0004-637X/831/2/187}

\bibitem[{{Arzoumanian} {et~al.}(2002){Arzoumanian}, {Chernoff}, \&
  {Cordes}}]{arzoumanian02}
{Arzoumanian}, Z., {Chernoff}, D.~F., \& {Cordes}, J.~M. 2002, \apj, 568, 289,
  \dodoi{10.1086/338805}

\bibitem[{{Askar} {et~al.}(2017){Askar}, {Szkudlarek}, {Gondek-Rosi\'{n}ska},
  {Giersz}, \& {Bulik}}]{askar17}
{Askar}, A., {Szkudlarek}, M., {Gondek-Rosi\'{n}ska}, D., {Giersz}, M., \&
  {Bulik}, T. 2017, \mnras, 464, L36, \dodoi{10.1093/mnrasl/slw177}

\bibitem[{{Banerjee}(2017)}]{banerjee17}
{Banerjee}, S. 2017, \mnras, 467, 524, \dodoi{10.1093/mnras/stw3392}

\bibitem[{{Banerjee}(2018{\natexlab{a}})}]{banerjee18}
---. 2018{\natexlab{a}}, \mnras, 473, 909, \dodoi{10.1093/mnras/stx2347}

\bibitem[{{Banerjee}(2018{\natexlab{b}})}]{banerjee18b}
---. 2018{\natexlab{b}}, \mnras, 481, 5123, \dodoi{10.1093/mnras/sty2608}

\bibitem[{{Banerjee} {et~al.}(2020){Banerjee}, {Belczynski}, {Fryer},
  {Berczik}, \& et~al.}]{banerjee20}
{Banerjee}, S., {Belczynski}, K., {Fryer}, C.~L., {Berczik}, P., \& et~al.
  2020, \aap, 639, A41, \dodoi{10.1051/0004-6361/201935332}

\bibitem[{{Bartos} {et~al.}(2017){Bartos}, {Kocsis}, {Haiman}, \&
  {M{\'a}rka}}]{bartos17}
{Bartos}, I., {Kocsis}, B., {Haiman}, Z., \& {M{\'a}rka}, S. 2017, \apj, 835,
  165, \dodoi{10.3847/1538-4357/835/2/165}

\bibitem[{{Bavera} {et~al.}(2021){Bavera}, {Fragos}, {Zevin}, {Berry},
  {Marchant}, {Andrews}, {Coughlin}, {Dotter}, {Kovlakas}, {Misra},
  {Serra-Perez}, {Qin}, {Rocha}, {Rom{\'a}n-Garza}, {Tran}, \&
  {Zapartas}}]{bavera21}
{Bavera}, S.~S., {Fragos}, T., {Zevin}, M., {et~al.} 2021, \aap, 647, A153,
  \dodoi{10.1051/0004-6361/202039804}

\bibitem[{{Belczynski} {et~al.}(2016{\natexlab{a}}){Belczynski}, {Holz},
  {Bulik}, \& {O'Shaughnessy}}]{belczynski16a}
{Belczynski}, K., {Holz}, D.~E., {Bulik}, T., \& {O'Shaughnessy}, R.
  2016{\natexlab{a}}, \nat, 534, 512, \dodoi{10.1038/nature18322}

\bibitem[{{Belczynski} {et~al.}(2002){Belczynski}, {Kalogera}, \&
  {Bulik}}]{belczynski02}
{Belczynski}, K., {Kalogera}, V., \& {Bulik}, T. 2002, \apj, 572, 407,
  \dodoi{10.1086/340304}

\bibitem[{{Belczynski} {et~al.}(2008){Belczynski}, {Kalogera}, {Rasio}, {Taam},
  {Zezas}, {Bulik}, {Maccarone}, \& {Ivanova}}]{belczynski08}
{Belczynski}, K., {Kalogera}, V., {Rasio}, F.~A., {et~al.} 2008, \apjs, 174,
  223, \dodoi{10.1086/521026}

\bibitem[{{Belczynski} {et~al.}(2016{\natexlab{b}}){Belczynski}, {Repetto},
  {Holz}, {O'Shaughnessy}, {Bulik}, {Berti}, {Fryer}, \&
  {Dominik}}]{belczynski16b}
{Belczynski}, K., {Repetto}, S., {Holz}, D.~E., {et~al.} 2016{\natexlab{b}},
  \apj, 819, 108, \dodoi{10.3847/0004-637X/819/2/108}

\bibitem[{{Belczynski} {et~al.}(2020){Belczynski}, {Klencki}, {Fields},
  {Olejak}, {Berti}, {Meynet}, {Fryer}, {Holz}, {O'Shaughnessy}, {Brown},
  {Bulik}, {Leung}, {Nomoto}, {Madau}, {Hirschi}, {Kaiser}, {Jones}, {Mondal},
  {Chruslinska}, {Drozda}, {Gerosa}, {Doctor}, {Giersz}, {Ekstrom}, {Georgy},
  {Askar}, {Baibhav}, {Wysocki}, {Natan}, {Farr}, {Wiktorowicz}, {Coleman
  Miller}, {Farr}, \& {Lasota}}]{belczynski20}
{Belczynski}, K., {Klencki}, J., {Fields}, C.~E., {et~al.} 2020, \aap, 636,
  A104, \dodoi{10.1051/0004-6361/201936528}

\bibitem[{{Bond} {et~al.}(1984){Bond}, {Arnett}, \& {Carr}}]{bond84}
{Bond}, J.~R., {Arnett}, W.~D., \& {Carr}, B.~J. 1984, \apj, 280, 825,
  \dodoi{10.1086/162057}

\bibitem[{{Brandt} \& {Podsiadlowski}(1995)}]{brandt95}
{Brandt}, N., \& {Podsiadlowski}, P. 1995, \mnras, 274, 461,
  \dodoi{10.1093/mnras/274.2.461}

\bibitem[{{Buonanno} {et~al.}(2008){Buonanno}, {Kidder}, \&
  {Lehner}}]{buonanno08}
{Buonanno}, A., {Kidder}, L.~E., \& {Lehner}, L. 2008, \prd, 77, 026004,
  \dodoi{10.1103/PhysRevD.77.026004}

\bibitem[{{Chandrasekhar}(1943)}]{chandrasekhar43}
{Chandrasekhar}, S. 1943, \apj, 97, 255, \dodoi{10.1086/144517}

\bibitem[{{Costa} {et~al.}(2021){Costa}, {Bressan}, {Mapelli}, {Marigo}, \&
  et~al.}]{costa21}
{Costa}, G., {Bressan}, A., {Mapelli}, M., {Marigo}, P., \& et~al. 2021,
  \mnras, 501, 4514, \dodoi{10.1093/mnras/staa3916}

\bibitem[{{Davis} {et~al.}(2022){Davis}, {Littenberg}, {Romero-Shaw},
  {Millhouse}, {McIver}, {Di Renzo}, \& {Ashton}}]{davis22}
{Davis}, D., {Littenberg}, T.~B., {Romero-Shaw}, I.~M., {et~al.} 2022,
  Classical and Quantum Gravity, 39, 245013, \dodoi{10.1088/1361-6382/aca238}

\bibitem[{{de Mink} {et~al.}(2009){de Mink}, {Cantiello}, {Langer}, {Pols},
  {Brott}, \& {Yoon}}]{demink09}
{de Mink}, S.~E., {Cantiello}, M., {Langer}, N., {et~al.} 2009, \aap, 497, 243,
  \dodoi{10.1051/0004-6361/200811439}

\bibitem[{{de Mink} \& {Mandel}(2016)}]{demink16}
{de Mink}, S.~E., \& {Mandel}, I. 2016, \mnras, 460, 3545,
  \dodoi{10.1093/mnras/stw1219}

\bibitem[{{Di Carlo} {et~al.}(2019){Di Carlo}, {Giacobbo}, {Mapelli},
  {Pasquato}, {Spera}, {Wang}, \& {Haardt}}]{dicarlo19}
{Di Carlo}, U.~N., {Giacobbo}, N., {Mapelli}, M., {et~al.} 2019, \mnras, 487,
  2947, \dodoi{10.1093/mnras/stz1453}

\bibitem[{{Dominik} {et~al.}(2012){Dominik}, {Belczynski}, {Fryer}, {Holz},
  {Berti}, {Bulik}, {Mandel}, \& {O'Shaughnessy}}]{dominik12}
{Dominik}, M., {Belczynski}, K., {Fryer}, C., {et~al.} 2012, \apj, 759, 52,
  \dodoi{10.1088/0004-637X/759/1/52}

\bibitem[{{Dominik} {et~al.}(2013){Dominik}, {Belczynski}, {Fryer}, {Holz}, \&
  et~al.}]{dominik13}
{Dominik}, M., {Belczynski}, K., {Fryer}, C., {Holz}, D.~E., \& et~al. 2013,
  \apj, 779, 72, \dodoi{10.1088/0004-637X/779/1/72}

\bibitem[{{eLISA Consortium} {et~al.}(2013){eLISA Consortium}, {Amaro Seoane},
  {Aoudia}, {Audley}, {Auger}, {Babak}, {Baker}, {Barausse}, {Barke}, {Bassan},
  {Beckmann}, {Benacquista}, {Bender}, {Berti}, {Bin{\'e}truy}, {Bogenstahl},
  {Bonvin}, {Bortoluzzi}, {Brause}, {Brossard}, {Buchman}, {Bykov}, {Camp},
  {Caprini}, {Cavalleri}, {Cerdonio}, {Ciani}, {Colpi}, {Congedo}, {Conklin},
  {Cornish}, {Danzmann}, {de Vine}, {DeBra}, {Dewi Freitag}, {Di Fiore}, {Diaz
  Aguilo}, {Diepholz}, {Dolesi}, {Dotti}, {Fern{\'a}ndez Barranco},
  {Ferraioli}, {Ferroni}, {Finetti}, {Fitzsimons}, {Gair}, {Galeazzi},
  {Garcia}, {Gerberding}, {Gesa}, {Giardini}, {Gibert}, {Grimani}, {Groot},
  {Guzman Cervantes}, {Haiman}, {Halloin}, {Heinzel}, {Hewitson}, {Hogan},
  {Holz}, {Hornstrup}, {Hoyland}, {Hoyle}, {Hueller}, {Hughes}, {Jetzer},
  {Kalogera}, {Karnesis}, {Kilic}, {Killow}, {Klipstein}, {Kochkina},
  {Korsakova}, {Krolak}, {Larson}, {Lieser}, {Littenberg}, {Livas}, {Lloro},
  {Mance}, {Madau}, {Maghami}, {Mahrdt}, {Marsh}, {Mateos}, {Mayer},
  {McClelland}, {McKenzie}, {McWilliams}, {Merkowitz}, {Miller}, {Mitryk},
  {Moerschell}, {Mohanty}, {Monsky}, {Mueller}, {M{\"u}ller}, {Nelemans},
  {Nicolodi}, {Nissanke}, {Nofrarias}, {Numata}, {Ohme}, {Otto},
  {Perreur-Lloyd}, {Petiteau}, {Phinney}, {Plagnol}, {Pollack}, {Porter},
  {Prat}, {Preston}, {Prince}, {Reiche}, {Richstone}, {Robertson}, {Rossi},
  {Rosswog}, {Rubbo}, {Ruiter}, {Sanjuan}, {Sathyaprakash}, {Schlamminger},
  {Schutz}, {Sch{\"u}tze}, {Sesana}, {Shaddock}, {Shah}, {Sheard}, {Sopuerta},
  {Spector}, {Spero}, {Stanga}, {Stebbins}, {Stede}, {Steier}, {Sumner}, {Sun},
  {Sutton}, {Tanaka}, {Tanner}, {Thorpe}, {Tr{\"o}bs}, {Tinto}, {Tu},
  {Vallisneri}, {Vetrugno}, {Vitale}, {Volonteri}, {Wand}, {Wang}, {Wanner},
  {Ward}, {Ware}, {Wass}, {Weber}, {Yu}, {Yunes}, \& {Zweifel}}]{lisa13}
{eLISA Consortium}, {Amaro Seoane}, P., {Aoudia}, S., {et~al.} 2013, arXiv
  e-prints, arXiv:1305.5720.
\newblock \doarXiv{1305.5720}

\bibitem[{{Farmer} {et~al.}(2020){Farmer}, {Renzo}, {de Mink}, {Fishbach}, \&
  et~al.}]{farmer20}
{Farmer}, R., {Renzo}, M., {de Mink}, S.~E., {Fishbach}, M., \& et~al. 2020,
  \apjl, 902, L36, \dodoi{10.3847/2041-8213/abbadd}

\bibitem[{{Farmer} {et~al.}(2019){Farmer}, {Renzo}, {de Mink}, {Marchant}, \&
  {Justham}}]{farmer19}
{Farmer}, R., {Renzo}, M., {de Mink}, S.~E., {Marchant}, P., \& {Justham}, S.
  2019, \apj, 887, 53, \dodoi{10.3847/1538-4357/ab518b}

\bibitem[{{Fowler} \& {Hoyle}(1964)}]{fowler64}
{Fowler}, W.~A., \& {Hoyle}, F. 1964, \apjs, 9, 201, \dodoi{10.1086/190103}

\bibitem[{{Fragione} \& {Banerjee}(2021)}]{fragione21c}
{Fragione}, G., \& {Banerjee}, S. 2021, \apjl, 913, L29,
  \dodoi{10.3847/2041-8213/ac00a7}

\bibitem[{{Fragione} {et~al.}(2018){Fragione}, {Ginsburg}, \&
  {Kocsis}}]{fragione18b}
{Fragione}, G., {Ginsburg}, I., \& {Kocsis}, B. 2018, \apj, 856, 92,
  \dodoi{10.3847/1538-4357/aab368}

\bibitem[{{Fragione} {et~al.}(2019){Fragione}, {Grishin}, {Leigh}, {Perets}, \&
  et~al.}]{fragione19}
{Fragione}, G., {Grishin}, E., {Leigh}, N. W.~C., {Perets}, H.~B., \& et~al.
  2019, \mnras, 488, 47, \dodoi{10.1093/mnras/stz1651}

\bibitem[{{Fragione} \& {Kocsis}(2018)}]{fragione18}
{Fragione}, G., \& {Kocsis}, B. 2018, \prl, 121, 161103,
  \dodoi{10.1103/PhysRevLett.121.161103}

\bibitem[{{Fragione} {et~al.}(2022{\natexlab{a}}){Fragione}, {Kocsis}, {Rasio},
  \& {Silk}}]{fragione22}
{Fragione}, G., {Kocsis}, B., {Rasio}, F.~A., \& {Silk}, J. 2022{\natexlab{a}},
  \apj, 927, 231, \dodoi{10.3847/1538-4357/ac5026}

\bibitem[{{Fragione} {et~al.}(2022{\natexlab{b}}){Fragione}, {Loeb}, {Kocsis},
  \& {Rasio}}]{fragione22b}
{Fragione}, G., {Loeb}, A., {Kocsis}, B., \& {Rasio}, F.~A. 2022{\natexlab{b}},
  \apj, 933, 170, \dodoi{10.3847/1538-4357/ac75d0}

\bibitem[{{Fragione} {et~al.}(2020){Fragione}, {Loeb}, \&
  {Rasio}}]{fragione20c}
{Fragione}, G., {Loeb}, A., \& {Rasio}, F.~A. 2020, \apjl, 902, L26,
  \dodoi{10.3847/2041-8213/abbc0a}

\bibitem[{{Fragione} {et~al.}(2021){Fragione}, {Loeb}, \& {Rasio}}]{fragione21}
---. 2021, \apjl, 918, L38, \dodoi{10.3847/2041-8213/ac225a}

\bibitem[{{Fragione} \& {Silk}(2020)}]{fragione20b}
{Fragione}, G., \& {Silk}, J. 2020, \mnras, 498, 4591,
  \dodoi{10.1093/mnras/staa2629}

\bibitem[{{Fuller} \& {Ma}(2019)}]{fuller19}
{Fuller}, J., \& {Ma}, L. 2019, \apjl, 881, L1,
  \dodoi{10.3847/2041-8213/ab339b}

\bibitem[{{Galaudage} {et~al.}(2021){Galaudage}, {Talbot}, {Nagar}, {Jain},
  {Thrane}, \& {Mandel}}]{galaudage21}
{Galaudage}, S., {Talbot}, C., {Nagar}, T., {et~al.} 2021, \apjl, 921, L15,
  \dodoi{10.3847/2041-8213/ac2f3c}

\bibitem[{{Gayathri} {et~al.}(2022){Gayathri}, {Healy}, {Lange}, {O'Brien},
  {Szczepa{\'n}czyk}, {Bartos}, {Campanelli}, {Klimenko}, {Lousto}, \&
  {O'Shaughnessy}}]{gayathri22}
{Gayathri}, V., {Healy}, J., {Lange}, J., {et~al.} 2022, Nature Astronomy, 6,
  344, \dodoi{10.1038/s41550-021-01568-w}

\bibitem[{{Gerosa} \& {Berti}(2017)}]{gerosa17}
{Gerosa}, D., \& {Berti}, E. 2017, \prd, 95, 124046,
  \dodoi{10.1103/PhysRevD.95.124046}

\bibitem[{{Giacobbo} \& {Mapelli}(2018)}]{giacobbo18}
{Giacobbo}, N., \& {Mapelli}, M. 2018, \mnras, 480, 2011,
  \dodoi{10.1093/mnras/sty1999}

\bibitem[{{Gonz{\'a}lez} {et~al.}(2021){Gonz{\'a}lez}, {Kremer}, {Chatterjee},
  {Fragione}, \& et~al.}]{gonzalez21}
{Gonz{\'a}lez}, E., {Kremer}, K., {Chatterjee}, S., {Fragione}, G., \& et~al.
  2021, \apjl, 908, L29, \dodoi{10.3847/2041-8213/abdf5b}

\bibitem[{{Hamers} {et~al.}(2021){Hamers}, {Fragione}, {Neunteufel}, \&
  {Kocsis}}]{hamers21}
{Hamers}, A.~S., {Fragione}, G., {Neunteufel}, P., \& {Kocsis}, B. 2021,
  \mnras, 506, 5345, \dodoi{10.1093/mnras/stab2136}

\bibitem[{{Heger} \& {Woosley}(2002)}]{heger02}
{Heger}, A., \& {Woosley}, S.~E. 2002, \apj, 567, 532, \dodoi{10.1086/338487}

\bibitem[{Heggie(1975)}]{heggie75}
Heggie, D.~C. 1975, Mon.~Not.~R.~Astron.~Soc, 173, 729.
\newblock \url{http://adsabs.harvard.edu/abs/1975MNRAS.173..729H
  papers3://publication/uuid/EEE9B361-0082-4772-8BB8-23C720F9704E}

\bibitem[{{Hills}(1983)}]{hills83}
{Hills}, J.~G. 1983, \apj, 267, 322, \dodoi{10.1086/160871}

\bibitem[{{Hoang} {et~al.}(2018){Hoang}, {Naoz}, {Kocsis}, {Rasio}, \&
  et~al.}]{hoang18}
{Hoang}, B.-M., {Naoz}, S., {Kocsis}, B., {Rasio}, F.~A., \& et~al. 2018, \apj,
  856, 140, \dodoi{10.3847/1538-4357/aaafce}

\bibitem[{{Hobbs} {et~al.}(2005){Hobbs}, {Lorimer}, {Lyne}, \&
  {Kramer}}]{hobbs05}
{Hobbs}, G., {Lorimer}, D.~R., {Lyne}, A.~G., \& {Kramer}, M. 2005, \mnras,
  360, 974, \dodoi{10.1111/j.1365-2966.2005.09087.x}

\bibitem[{Hoy \& Raymond(2021)}]{hoy20}
Hoy, C., \& Raymond, V. 2021, SoftwareX, 15, 100765,
  \dodoi{10.1016/j.softx.2021.100765}

\bibitem[{{Hurley} {et~al.}(2000){Hurley}, {Pols}, \& {Tout}}]{hurley00}
{Hurley}, J.~R., {Pols}, O.~R., \& {Tout}, C.~A. 2000, \mnras, 315, 543,
  \dodoi{10.1046/j.1365-8711.2000.03426.x}

\bibitem[{{Hut} {et~al.}(1992){Hut}, {McMillan}, {Goodman}, {Mateo}, {Phinney},
  {Pryor}, {Richer}, {Verbunt}, \& {Weinberg}}]{hut92}
{Hut}, P., {McMillan}, S., {Goodman}, J., {et~al.} 1992, \pasp, 104, 981,
  \dodoi{10.1086/133085}

\bibitem[{{Janka}(2013)}]{janka13}
{Janka}, H.-T. 2013, \mnras, 434, 1355, \dodoi{10.1093/mnras/stt1106}

\bibitem[{{Kalogera}(1996)}]{kalogera96}
{Kalogera}, V. 1996, \apj, 471, 352, \dodoi{10.1086/177974}

\bibitem[{{Kalogera}(2000)}]{kalogera00}
---. 2000, \apj, 541, 319, \dodoi{10.1086/309400}

\bibitem[{{Kalogera} {et~al.}(2007){Kalogera}, {Belczynski}, {Kim},
  {O'Shaughnessy}, \& {Willems}}]{kalogera07}
{Kalogera}, V., {Belczynski}, K., {Kim}, C., {O'Shaughnessy}, R., \& {Willems},
  B. 2007, \physrep, 442, 75, \dodoi{10.1016/j.physrep.2007.02.008}

\bibitem[{{Kapil} {et~al.}(2023){Kapil}, {Mandel}, {Berti}, \&
  {M{\"u}ller}}]{kapil23}
{Kapil}, V., {Mandel}, I., {Berti}, E., \& {M{\"u}ller}, B. 2023, \mnras, 519,
  5893, \dodoi{10.1093/mnras/stad019}

\bibitem[{Kim {et~al.}(2003)Kim, Kalogera, \& Lorimer}]{kim03}
Kim, C., Kalogera, V., \& Lorimer, D.~R. 2003, Astrophys. J., 584, 985,
  \dodoi{10.1086/345740}

\bibitem[{{Kimball} {et~al.}(2021){Kimball}, {Talbot}, {Berry}, {Zevin}, \&
  et~al.}]{kimball21}
{Kimball}, C., {Talbot}, C., {Berry}, C. P.~L., {Zevin}, M., \& et~al. 2021,
  \apjl, 915, L35, \dodoi{10.3847/2041-8213/ac0aef}

\bibitem[{{Kremer} {et~al.}(2020){Kremer}, {Ye}, {Rui}, {Weatherford},
  {Chatterjee}, {Fragione}, {Rodriguez}, {Spera}, \& {Rasio}}]{kremer20}
{Kremer}, K., {Ye}, C.~S., {Rui}, N.~Z., {et~al.} 2020, \apjs, 247, 48,
  \dodoi{10.3847/1538-4365/ab7919}

\bibitem[{{Lousto} {et~al.}(2010){Lousto}, {Campanelli}, {Zlochower}, \&
  {Nakano}}]{lousto10}
{Lousto}, C.~O., {Campanelli}, M., {Zlochower}, Y., \& {Nakano}, H. 2010,
  Classical and Quantum Gravity, 27, 114006,
  \dodoi{10.1088/0264-9381/27/11/114006}

\bibitem[{{Lousto} {et~al.}(2012){Lousto}, {Zlochower}, {Dotti}, \&
  {Volonteri}}]{lousto12}
{Lousto}, C.~O., {Zlochower}, Y., {Dotti}, M., \& {Volonteri}, M. 2012, \prd,
  85, 084015, \dodoi{10.1103/PhysRevD.85.084015}

\bibitem[{{Maggiore} {et~al.}(2020){Maggiore}, {Van Den Broeck}, {Bartolo},
  {Belgacem}, {Bertacca}, {Bizouard}, {Branchesi}, {Clesse}, {Foffa},
  {Garc{\'\i}a-Bellido}, {Grimm}, {Harms}, {Hinderer}, {Matarrese}, {Palomba},
  {Peloso}, {Ricciardone}, \& {Sakellariadou}}]{maggiore20}
{Maggiore}, M., {Van Den Broeck}, C., {Bartolo}, N., {et~al.} 2020, \jcap,
  2020, 050, \dodoi{10.1088/1475-7516/2020/03/050}

\bibitem[{{Mandel} \& {de Mink}(2016)}]{mandel16}
{Mandel}, I., \& {de Mink}, S.~E. 2016, \mnras, 458, 2634,
  \dodoi{10.1093/mnras/stw379}

\bibitem[{{Mapelli} {et~al.}(2021){Mapelli}, {Dall'Amico}, {Bouffanais},
  {Giacobbo}, {Arca Sedda}, {Artale}, {Ballone}, {Di Carlo}, {Iorio},
  {Santoliquido}, \& {Torniamenti}}]{mapelli21}
{Mapelli}, M., {Dall'Amico}, M., {Bouffanais}, Y., {et~al.} 2021, \mnras, 505,
  339, \dodoi{10.1093/mnras/stab1334}

\bibitem[{{Marchant} {et~al.}(2016){Marchant}, {Langer}, {Podsiadlowski},
  {Tauris}, \& {Moriya}}]{marchant16}
{Marchant}, P., {Langer}, N., {Podsiadlowski}, P., {Tauris}, T.~M., \&
  {Moriya}, T.~J. 2016, \aap, 588, A50, \dodoi{10.1051/0004-6361/201628133}

\bibitem[{{Martinez} {et~al.}(2022){Martinez}, {Rodriguez}, \&
  {Fragione}}]{martinez22}
{Martinez}, M. A.~S., {Rodriguez}, C.~L., \& {Fragione}, G. 2022, \apj, 937,
  78, \dodoi{10.3847/1538-4357/ac8d55}

\bibitem[{{Martinez} {et~al.}(2020){Martinez}, {Fragione}, {Kremer},
  {Chatterjee}, {Rodriguez}, {Samsing}, {Ye}, {Weatherford}, {Zevin}, {Naoz},
  \& {Rasio}}]{martinez20}
{Martinez}, M. A.~S., {Fragione}, G., {Kremer}, K., {et~al.} 2020, \apj, 903,
  67, \dodoi{10.3847/1538-4357/abba25}

\bibitem[{{McMillan} {et~al.}(1991){McMillan}, {Hut}, \& {Makino}}]{mcmillan91}
{McMillan}, S., {Hut}, P., \& {Makino}, J. 1991, \apj, 372, 111,
  \dodoi{10.1086/169958}

\bibitem[{{Miller} \& {Hamilton}(2002)}]{miller02b}
{Miller}, M.~C., \& {Hamilton}, D.~P. 2002, \apj, 576, 894,
  \dodoi{10.1086/341788}

\bibitem[{{Neijssel} {et~al.}(2019){Neijssel}, {Vigna-G{\'o}mez}, {Stevenson},
  {Barrett}, {Gaebel}, {Broekgaarden}, {de Mink}, {Sz{\'e}csi}, {Vinciguerra},
  \& {Mandel}}]{neijssel19}
{Neijssel}, C.~J., {Vigna-G{\'o}mez}, A., {Stevenson}, S., {et~al.} 2019,
  \mnras, 490, 3740, \dodoi{10.1093/mnras/stz2840}

\bibitem[{{Paczynski}(1976)}]{paczynski76}
{Paczynski}, B. 1976, in Structure and Evolution of Close Binary Systems, ed.
  P.~{Eggleton}, S.~{Mitton}, \& J.~{Whelan}, Vol.~73, 75

\bibitem[{{Payne} {et~al.}(2019){Payne}, {Talbot}, \& {Thrane}}]{payne19}
{Payne}, E., {Talbot}, C., \& {Thrane}, E. 2019, \prd, 100, 123017,
  \dodoi{10.1103/PhysRevD.100.123017}

\bibitem[{{Peters}(1964)}]{peters64}
{Peters}, P.~C. 1964, Physical Review, 136, 1224,
  \dodoi{10.1103/PhysRev.136.B1224}

\bibitem[{{Pijloo} {et~al.}(2012){Pijloo}, {Caputo}, \& {Portegies
  Zwart}}]{pijloo12}
{Pijloo}, J.~T., {Caputo}, D.~P., \& {Portegies Zwart}, S.~F. 2012, \mnras,
  424, 2914, \dodoi{10.1111/j.1365-2966.2012.21431.x}

\bibitem[{{Portegies Zwart} \& {McMillan}(2000)}]{portegies00}
{Portegies Zwart}, S.~F., \& {McMillan}, S. L.~W. 2000, \apjl, 528, L17,
  \dodoi{10.1086/312422}

\bibitem[{{Postnov} \& {Yungelson}(2014)}]{postnov14}
{Postnov}, K.~A., \& {Yungelson}, L.~R. 2014, Living Reviews in Relativity, 17,
  3, \dodoi{10.12942/lrr-2014-3}

\bibitem[{{Qin} {et~al.}(2018){Qin}, {Fragos}, {Meynet}, {Andrews},
  {S{\o}rensen}, \& {Song}}]{qin18}
{Qin}, Y., {Fragos}, T., {Meynet}, G., {et~al.} 2018, \aap, 616, A28,
  \dodoi{10.1051/0004-6361/201832839}

\bibitem[{{Reitze} {et~al.}(2019){Reitze}, {Adhikari}, {Ballmer}, {Barish},
  {Barsotti}, {Billingsley}, {Brown}, {Chen}, {Coyne}, {Eisenstein}, {Evans},
  {Fritschel}, {Hall}, {Lazzarini}, {Lovelace}, {Read}, {Sathyaprakash},
  {Shoemaker}, {Smith}, {Torrie}, {Vitale}, {Weiss}, {Wipf}, \&
  {Zucker}}]{reitze19}
{Reitze}, D., {Adhikari}, R.~X., {Ballmer}, S., {et~al.} 2019, in Bulletin of
  the American Astronomical Society, Vol.~51, 35.
\newblock \doarXiv{1907.04833}

\bibitem[{{Renzo} {et~al.}(2020){Renzo}, {Farmer}, {Justham}, {de Mink}, \&
  et~al.}]{renzo20b}
{Renzo}, M., {Farmer}, R.~J., {Justham}, S., {de Mink}, S.~E., \& et~al. 2020,
  \mnras, 493, 4333, \dodoi{10.1093/mnras/staa549}

\bibitem[{{Repetto} {et~al.}(2012){Repetto}, {Davies}, \&
  {Sigurdsson}}]{repetto12}
{Repetto}, S., {Davies}, M.~B., \& {Sigurdsson}, S. 2012, \mnras, 425, 2799,
  \dodoi{10.1111/j.1365-2966.2012.21549.x}

\bibitem[{{Rodriguez} {et~al.}(2018){Rodriguez}, {Amaro-Seoane}, {Chatterjee},
  \& {Rasio}}]{rodriguez18}
{Rodriguez}, C.~L., {Amaro-Seoane}, P., {Chatterjee}, S., \& {Rasio}, F.~A.
  2018, PRL, 120, 151101, \dodoi{10.1103/PhysRevLett.120.151101}

\bibitem[{{Rodriguez} {et~al.}(2016){Rodriguez}, {Chatterjee}, \&
  {Rasio}}]{rodriguez16b}
{Rodriguez}, C.~L., {Chatterjee}, S., \& {Rasio}, F.~A. 2016, \prd, 93, 084029,
  \dodoi{10.1103/PhysRevD.93.084029}

\bibitem[{{Rodriguez} {et~al.}(2020){Rodriguez}, {Kremer}, {Grudi{\'c}},
  {Hafen}, \& et~al.}]{rodriguez20}
{Rodriguez}, C.~L., {Kremer}, K., {Grudi{\'c}}, M.~Y., {Hafen}, Z., \& et~al.
  2020, \apjl, 896, L10, \dodoi{10.3847/2041-8213/ab961d}

\bibitem[{{Rodriguez} {et~al.}(2015){Rodriguez}, {Morscher}, {Pattabiraman},
  {Chatterjee}, {Haster}, \& {Rasio}}]{rodriguez15}
{Rodriguez}, C.~L., {Morscher}, M., {Pattabiraman}, B., {et~al.} 2015, \prl,
  115, 051101, \dodoi{10.1103/PhysRevLett.115.051101}

\bibitem[{{Rodriguez} {et~al.}(2019){Rodriguez}, {Zevin}, {Amaro-Seoane},
  {Chatterjee}, {Kremer}, {Rasio}, \& {Ye}}]{rodriguez19}
{Rodriguez}, C.~L., {Zevin}, M., {Amaro-Seoane}, P., {et~al.} 2019, \prd, 100,
  043027, \dodoi{10.1103/PhysRevD.100.043027}

\bibitem[{{Romero-Shaw} {et~al.}(2019){Romero-Shaw}, {Lasky}, \&
  {Thrane}}]{romero-shaw19}
{Romero-Shaw}, I.~M., {Lasky}, P.~D., \& {Thrane}, E. 2019, \mnras, 490, 5210,
  \dodoi{10.1093/mnras/stz2996}

\bibitem[{{Romero-Shaw} {et~al.}(2022){Romero-Shaw}, {Lasky}, \&
  {Thrane}}]{romero-shaw22}
---. 2022, arXiv e-prints, arXiv:2206.14695.
\newblock \doarXiv{2206.14695}

\bibitem[{{Roulet} {et~al.}(2021){Roulet}, {Chia}, {Olsen}, {Dai},
  {Venumadhav}, {Zackay}, \& {Zaldarriaga}}]{roulet21}
{Roulet}, J., {Chia}, H.~S., {Olsen}, S., {et~al.} 2021, \prd, 104, 083010,
  \dodoi{10.1103/PhysRevD.104.083010}

\bibitem[{{Samsing}(2018)}]{samsing18b}
{Samsing}, J. 2018, \prd, 97, 103014, \dodoi{10.1103/PhysRevD.97.103014}

\bibitem[{{Samsing} \& {D'Orazio}(2018)}]{samsing18c}
{Samsing}, J., \& {D'Orazio}, D.~J. 2018, \mnras, 481, 5445,
  \dodoi{10.1093/mnras/sty2334}

\bibitem[{{Samsing} {et~al.}(2018){Samsing}, {MacLeod}, \&
  {Ramirez-Ruiz}}]{samsing18d}
{Samsing}, J., {MacLeod}, M., \& {Ramirez-Ruiz}, E. 2018, \apj, 853, 140,
  \dodoi{10.3847/1538-4357/aaa715}

\bibitem[{{Spera} {et~al.}(2019){Spera}, {Mapelli}, {Giacobbo}, {Trani},
  {Bressan}, \& {Costa}}]{spera19}
{Spera}, M., {Mapelli}, M., {Giacobbo}, N., {et~al.} 2019, \mnras, 485, 889,
  \dodoi{10.1093/mnras/stz359}

\bibitem[{{Spitzer}(1987)}]{spitzer87}
{Spitzer}, L. 1987, {Dynamical evolution of globular clusters} (Princeton
  University Press)

\bibitem[{{Stevenson} {et~al.}(2017){Stevenson}, {Vigna-G{\'o}mez}, {Mandel},
  {Barrett}, {Neijssel}, {Perkins}, \& {de Mink}}]{stevenson17}
{Stevenson}, S., {Vigna-G{\'o}mez}, A., {Mandel}, I., {et~al.} 2017, Nature
  Communications, 8, 14906, \dodoi{10.1038/ncomms14906}

\bibitem[{{Su} {et~al.}(2021){Su}, {Liu}, \& {Lai}}]{su21}
{Su}, Y., {Liu}, B., \& {Lai}, D. 2021, \mnras, 505, 3681,
  \dodoi{10.1093/mnras/stab1617}

\bibitem[{{Tagawa} {et~al.}(2020{\natexlab{a}}){Tagawa}, {Haiman}, {Bartos}, \&
  {Kocsis}}]{tagawa20c}
{Tagawa}, H., {Haiman}, Z., {Bartos}, I., \& {Kocsis}, B. 2020{\natexlab{a}},
  \apj, 899, 26, \dodoi{10.3847/1538-4357/aba2cc}

\bibitem[{{Tagawa} {et~al.}(2020{\natexlab{b}}){Tagawa}, {Haiman}, \&
  {Kocsis}}]{tagawa20}
{Tagawa}, H., {Haiman}, Z., \& {Kocsis}, B. 2020{\natexlab{b}}, \apj, 898, 25,
  \dodoi{10.3847/1538-4357/ab9b8c}

\bibitem[{{Tagawa} {et~al.}(2020{\natexlab{c}}){Tagawa}, {Haiman}, \&
  {Kocsis}}]{tagawa20b}
---. 2020{\natexlab{c}}, \apj, 892, 36, \dodoi{10.3847/1538-4357/ab7922}

\bibitem[{{Tagawa} {et~al.}(2018){Tagawa}, {Saitoh}, \& {Kocsis}}]{tagawa18}
{Tagawa}, H., {Saitoh}, T.~R., \& {Kocsis}, B. 2018, \prl, 120, 261101,
  \dodoi{10.1103/PhysRevLett.120.261101}

\bibitem[{{Tauris}(2022)}]{tauris22}
{Tauris}, T.~M. 2022, \apj, 938, 66, \dodoi{10.3847/1538-4357/ac86c8}

\bibitem[{{The LIGO Scientific Collaboration} {et~al.}(2021){The LIGO
  Scientific Collaboration}, {the Virgo Collaboration}, {the KAGRA
  Collaboration}, {Abbott}, {Abbott}, {Acernese}, {Ackley}, {Adams},
  {Adhikari}, {Adhikari}, {Adya}, {Affeldt}, {Agarwal}, {Agathos}, {Agatsuma},
  {Aggarwal}, {Aguiar}, {Aiello}, {Ain}, {Ajith}, {Akcay}, {Akutsu},
  {Albanesi}, {Allocca}, {Altin}, {Amato}, {Anand}, {Anand}, {Ananyeva},
  {Anderson}, {Anderson}, {Ando}, {Andrade}, {Andres}, {Andri{\'c}},
  {Angelova}, {Ansoldi}, {Antelis}, {Antier}, {Appert}, {Arai}, {Arai}, {Arai},
  {Araki}, {Araya}, {Araya}, {Areeda}, {Ar{\`e}ne}, {Aritomi}, {Arnaud},
  {Arogeti}, {Aronson}, {Arun}, {Asada}, {Asali}, {Ashton}, {Aso}, {Assiduo},
  {Aston}, {Astone}, {Aubin}, {Austin}, {Babak}, {Badaracco}, {Bader},
  {Badger}, {Bae}, {Bae}, {Baer}, {Bagnasco}, {Bai}, {Baiotti}, {Baird},
  {Bajpai}, {Ball}, {Ballardin}, {Ballmer}, {Balsamo}, {Baltus}, {Banagiri},
  {Bankar}, {Barayoga}, {Barbieri}, {Barish}, {Barker}, {Barneo}, {Barone},
  {Barr}, {Barsotti}, {Barsuglia}, {Barta}, {Bartlett}, {Barton}, {Bartos},
  {Bassiri}, {Basti}, {Bawaj}, {Bayley}, {Baylor}, {Bazzan}, {B{\'e}csy},
  {Bedakihale}, {Bejger}, {Belahcene}, {Benedetto}, {Beniwal}, {Bennett},
  {Bentley}, {BenYaala}, {Bergamin}, {Berger}, {Bernuzzi}, {Berry},
  {Bersanetti}, {Bertolini}, {Betzwieser}, {Beveridge}, {Bhandare}, {Bhardwaj},
  {Bhattacharjee}, {Bhaumik}, {Bilenko}, {Billingsley}, {Bini}, {Birney},
  {Birnholtz}, {Biscans}, {Bischi}, {Biscoveanu}, {Bisht}, {Biswas}, {Bitossi},
  {Bizouard}, {Blackburn}, {Blair}, {Blair}, {Blair}, {Bobba}, {Bode}, {Boer},
  {Bogaert}, {Boldrini}, {Bonavena}, {Bondu}, {Bonilla}, {Bonnand}, {Booker},
  {Boom}, {Bork}, {Boschi}, {Bose}, {Bose}, {Bossilkov}, {Boudart},
  {Bouffanais}, {Bozzi}, {Bradaschia}, {Brady}, {Bramley}, {Branch},
  {Branchesi}, {Brandt}, {Brau}, {Breschi}, {Briant}, {Briggs}, {Brillet},
  {Brinkmann}, {Brockill}, {Brooks}, {Brooks}, {Brown}, {Brunett}, {Bruno},
  {Bruntz}, {Bryant}, {Bulik}, {Bulten}, {Buonanno}, {Buscicchio}, {Buskulic},
  {Buy}, {Byer}, {Cabourn Davies}, {Cadonati}, {Cagnoli}, {Cahillane},
  {Calder{\'o}n Bustillo}, {Callaghan}, {Callister}, {Calloni}, {Cameron},
  {Camp}, {Canepa}, {Canevarolo}, {Cannavacciuolo}, {Cannon}, {Cao}, {Cao},
  {Capocasa}, {Capote}, {Carapella}, {Carbognani}, {Carlin}, {Carney},
  {Carpinelli}, {Carrillo}, {Carullo}, {Carver}, {Casanueva Diaz}, {Casentini},
  {Castaldi}, {Caudill}, {Cavagli{\`a}}, {Cavalier}, {Cavalieri}, {Ceasar},
  {Cella}, {Cerd{\'a}-Dur{\'a}n}, {Cesarini}, {Chaibi}, {Chakravarti},
  {Chalathadka Subrahmanya}, {Champion}, {Chan}, {Chan}, {Chan}, {Chan},
  {Chan}, {Chandra}, {Chanial}, {Chao}, {Chapman-Bird}, {Charlton}, {Chase},
  {Chassande-Mottin}, {Chatterjee}, {Chatterjee}, {Chatterjee}, {Chaturvedi},
  {Chaty}, {Chatziioannou}, {Chen}, {Chen}, {Chen}, {Chen}, {Chen}, {Chen},
  {Chen}, {Chen}, {Cheng}, {Cheong}, {Cheung}, {Chia}, {Chiadini}, {Chiang},
  {Chiarini}, {Chierici}, {Chincarini}, {Chiofalo}, {Chiummo}, {Cho}, {Cho},
  {Choudhary}, {Choudhary}, {Christensen}, {Chu}, {Chu}, {Chu}, {Chua},
  {Chung}, {Ciani}, {Ciecielag}, {Cie{\'s}lar}, {Cifaldi}, {Ciobanu}, {Ciolfi},
  {Cipriano}, {Cirone}, {Clara}, {Clark}, {Clark}, {Clarke}, {Clearwater},
  {Clesse}, {Cleva}, {Coccia}, {Codazzo}, {Cohadon}, {Cohen}, {Cohen},
  {Colleoni}, {Collette}, {Colombo}, {Colpi}, {Compton}, {Constancio}, {Conti},
  {Cooper}, {Corban}, {Corbitt}, {Cordero-Carri{\'o}n}, {Corezzi}, {Corley},
  {Cornish}, {Corre}, {Corsi}, {Cortese}, {Costa}, {Cotesta}, {Coughlin},
  {Coulon}, {Countryman}, {Cousins}, {Couvares}, {Coward}, {Cowart}, {Coyne},
  {Coyne}, {Creighton}, {Creighton}, {Criswell}, {Croquette}, {Crowder},
  {Cudell}, {Cullen}, {Cumming}, {Cummings}, {Cunningham}, {Cuoco},
  {Cury{\l}o}, {Dabadie}, {Dal Canton}, {Dall'Osso}, {D{\'a}lya}, {Dana},
  {DaneshgaranBajastani}, {D'Angelo}, {Danila}, {Danilishin}, {D'Antonio},
  {Danzmann}, {Darsow-Fromm}, {Dasgupta}, {Datrier}, {Datta}, {Dattilo},
  {Dave}, {Davier}, {Davis}, {Davis}, {Daw}, {de Alarc{\'o}n}, {Dean}, {DeBra},
  {Deenadayalan}, {Degallaix}, {De Laurentis}, {Del{\'e}glise}, {Del Favero},
  {De Lillo}, {De Lillo}, {Del Pozzo}, {DeMarchi}, {De Matteis}, {D'Emilio},
  {Demos}, {Dent}, {Depasse}, {De Pietri}, {De Rosa}, {De Rossi}, {DeSalvo},
  {De Simone}, {Dhurandhar}, {D{\'\i}az}, {Diaz-Ortiz}, {Didio}, {Dietrich},
  {Di Fiore}, {Di Fronzo}, {Di Giorgio}, {Di Giovanni}, {Di Giovanni}, {Di
  Girolamo}, {Di Lieto}, {Ding}, {Di Pace}, {Di Palma}, {Di Renzo},
  {Divakarla}, {Dmitriev}, {Doctor}, {D'Onofrio}, {Donovan}, {Dooley},
  {Doravari}, {Dorrington}, {Drago}, {Driggers}, {Drori}, {Ducoin}, {Dupej},
  {Durante}, {D'Urso}, {Duverne}, {Dwyer}, {Eassa}, {Easter}, {Ebersold},
  {Eckhardt}, {Eddolls}, {Edelman}, {Edo}, {Edy}, {Effler}, {Eguchi},
  {Eichholz}, {Eikenberry}, {Eisenmann}, {Eisenstein}, {Ejlli}, {Engelby},
  {Enomoto}, {Errico}, {Essick}, {Estell{\'e}s}, {Estevez}, {Etienne}, {Etzel},
  {Evans}, {Evans}, {Ewing}, {Fafone}, {Fair}, {Fairhurst}, {Farah}, {Farinon},
  {Farr}, {Farr}, {Farrow}, {Fauchon-Jones}, {Favaro}, {Favata}, {Fays},
  {Fazio}, {Feicht}, {Fejer}, {Fenyvesi}, {Ferguson}, {Fernandez-Galiana},
  {Ferrante}, {Ferreira}, {Fidecaro}, {Figura}, {Fiori}, {Fishbach}, {Fisher},
  {Fittipaldi}, {Fiumara}, {Flaminio}, {Floden}, {Fong}, {Font}, {Fornal},
  {Forsyth}, {Franke}, {Frasca}, {Frasconi}, {Frederick}, {Freed}, {Frei},
  {Freise}, {Frey}, {Fritschel}, {Frolov}, {Fronz{\'e}}, {Fujii}, {Fujikawa},
  {Fukunaga}, {Fukushima}, {Fulda}, {Fyffe}, {Gabbard}, {Gabella}, {Gadre},
  {Gair}, {Gais}, {Galaudage}, {Gamba}, {Ganapathy}, {Ganguly}, {Gao},
  {Gaonkar}, {Garaventa}, {Garc{\'\i}a}, {Garc{\'\i}a-N{\'u}{\~n}ez},
  {Garc{\'\i}a-Quir{\'o}s}, {Garufi}, {Gateley}, {Gaudio}, {Gayathri}, {Ge},
  {Gemme}, {Gennai}, {George}, {George}, {Gerberding}, {Gergely}, {Gewecke},
  {Ghonge}, {Ghosh}, {Ghosh}, {Ghosh}, {Ghosh}, {Giacomazzo}, {Giacoppo},
  {Giaime}, {Giardina}, {Gibson}, {Gier}, {Giesler}, {Giri}, {Gissi},
  {Glanzer}, {Gleckl}, {Godwin}, {Goetz}, {Goetz}, {Gohlke}, {Golomb},
  {Goncharov}, {Gonz{\'a}lez}, {Gopakumar}, {Gosselin}, {Gouaty}, {Gould},
  {Grace}, {Grado}, {Granata}, {Granata}, {Grant}, {Gras}, {Grassia}, {Gray},
  {Gray}, {Greco}, {Green}, {Green}, {Gretarsson}, {Gretarsson}, {Griffith},
  {Griffiths}, {Griggs}, {Grignani}, {Grimaldi}, {Grimm}, {Grote}, {Grunewald},
  {Gruning}, {Guerra}, {Guidi}, {Guimaraes}, {Guix{\'e}}, {Gulati}, {Guo},
  {Guo}, {Gupta}, {Gupta}, {Gupta}, {Gustafson}, {Gustafson}, {Guzman}, {Ha},
  {Haegel}, {Hagiwara}, {Haino}, {Halim}, {Hall}, {Hamilton}, {Hammond}, {Han},
  {Haney}, {Hanks}, {Hanna}, {Hannam}, {Hannuksela}, {Hansen}, {Hansen},
  {Hanson}, {Harder}, {Hardwick}, {Haris}, {Harms}, {Harry}, {Harry},
  {Hartwig}, {Hasegawa}, {Haskell}, {Hasskew}, {Haster}, {Hattori}, {Haughian},
  {Hayakawa}, {Hayama}, {Hayes}, {Healy}, {Heidmann}, {Heidt}, {Heintze},
  {Heinze}, {Heinzel}, {Heitmann}, {Hellman}, {Hello}, {Helmling-Cornell},
  {Hemming}, {Hendry}, {Heng}, {Hennes}, {Hennig}, {Hennig}, {Hernandez},
  {Hernandez Vivanco}, {Heurs}, {Hild}, {Hill}, {Himemoto}, {Hines},
  {Hiranuma}, {Hirata}, {Hirose}, {Hochheim}, {Hofman}, {Hohmann}, {Holcomb},
  {Holland}, {Holley-Bockelmann}, {Hollows}, {Holmes}, {Holt}, {Holz}, {Hong},
  {Hopkins}, {Hough}, {Hourihane}, {Howell}, {Hoy}, {Hoyland}, {Hreibi},
  {Hsieh}, {Hsu}, {Huang}, {Huang}, {Huang}, {Huang}, {Huang}, {Huang},
  {H{\"u}bner}, {Huddart}, {Hughey}, {Hui}, {Hui}, {Husa}, {Huttner},
  {Huxford}, {Huynh-Dinh}, {Ide}, {Idzkowski}, {Iess}, {Ikenoue}, {Imam},
  {Inayoshi}, {Ingram}, {Inoue}, {Ioka}, {Isi}, {Isleif}, {Ito}, {Itoh},
  {Iyer}, {Izumi}, {JaberianHamedan}, {Jacqmin}, {Jadhav}, {Jadhav}, {James},
  {Jan}, {Jani}, {Janquart}, {Janssens}, {Janthalur}, {Jaranowski}, {Jariwala},
  {Jaume}, {Jenkins}, {Jenner}, {Jeon}, {Jeunon}, {Jia}, {Jin}, {Johns},
  {Johnson-McDaniel}, {Jones}, {Jones}, {Jones}, {Jones}, {Jones}, {Jonker},
  {Ju}, {Jung}, {Jung}, {Junker}, {Juste}, {Kaihotsu}, {Kajita}, {Kakizaki},
  {Kalaghatgi}, {Kalogera}, {Kamai}, {Kamiizumi}, {Kanda}, {Kandhasamy},
  {Kang}, {Kanner}, {Kao}, {Kapadia}, {Kapasi}, {Karat}, {Karathanasis},
  {Karki}, {Kashyap}, {Kasprzack}, {Kastaun}, {Katsanevas}, {Katsavounidis},
  {Katzman}, {Kaur}, {Kawabe}, {Kawaguchi}, {Kawai}, {Kawasaki},
  {K{\'e}f{\'e}lian}, {Keitel}, {Key}, {Khadka}, {Khalili}, {Khan}, {Khazanov},
  {Khetan}, {Khursheed}, {Kijbunchoo}, {Kim}, {Kim}, {Kim}, {Kim}, {Kim},
  {Kim}, {Kimball}, {Kimura}, {Kinley-Hanlon}, {Kirchhoff}, {Kissel}, {Kita},
  {Kitazawa}, {Kleybolte}, {Klimenko}, {Knee}, {Knowles}, {Knyazev}, {Koch},
  {Koekoek}, {Kojima}, {Kokeyama}, {Koley}, {Kolitsidou}, {Kolstein}, {Komori},
  {Kondrashov}, {Kong}, {Kontos}, {Koper}, {Korobko}, {Kotake}, {Kovalam},
  {Kozak}, {Kozakai}, {Kozu}, {Kringel}, {Krishnendu}, {Kr{\'o}lak}, {Kuehn},
  {Kuei}, {Kuijer}, {Kulkarni}, {Kumar}, {Kumar}, {Kumar}, {Kumar}, {Kume},
  {Kuns}, {Kuo}, {Kuo}, {Kuromiya}, {Kuroyanagi}, {Kusayanagi}, {Kuwahara},
  {Kwak}, {Lagabbe}, {Laghi}, {Lalande}, {Lam}, {Lamberts}, {Landry}, {Lane},
  {Lang}, {Lange}, {Lantz}, {La Rosa}, {Lartaux-Vollard}, {Lasky}, {Laxen},
  {Lazzarini}, {Lazzaro}, {Leaci}, {Leavey}, {Lecoeuche}, {Lee}, {Lee}, {Lee},
  {Lee}, {Lee}, {Lee}, {Lehmann}, {Lema{\^\i}tre}, {Leonardi}, {Leroy},
  {Letendre}, {Levesque}, {Levin}, {Leviton}, {Leyde}, {Li}, {Li}, {Li}, {Li},
  {Li}, {Li}, {Lin}, {Lin}, {Lin}, {Lin}, {Lin}, {Linde}, {Linker}, {Linley},
  {Littenberg}, {Liu}, {Liu}, {Liu}, {Liu}, {Llamas}, {Llorens-Monteagudo},
  {Lo}, {Lockwood}, {Loh}, {London}, {Longo}, {Lopez}, {Lopez Portilla},
  {Lorenzini}, {Loriette}, {Lormand}, {Losurdo}, {Lott}, {Lough}, {Lousto},
  {Lovelace}, {Lucaccioni}, {L{\"u}ck}, {Lumaca}, {Lundgren}, {Luo}, {Lynam},
  {Macas}, {MacInnis}, {Macleod}, {MacMillan}, {Macquet}, {Maga{\~n}a
  Hernandez}, {Magazz{\`u}}, {Magee}, {Maggiore}, {Magnozzi}, {Mahesh},
  {Majorana}, {Makarem}, {Maksimovic}, {Maliakal}, {Malik}, {Man}, {Mandic},
  {Mangano}, {Mango}, {Mansell}, {Manske}, {Mantovani}, {Mapelli},
  {Marchesoni}, {Marchio}, {Marion}, {Mark}, {M{\'a}rka}, {M{\'a}rka},
  {Markakis}, {Markosyan}, {Markowitz}, {Maros}, {Marquina}, {Marsat},
  {Martelli}, {Martin}, {Martin}, {Martinez}, {Martinez}, {Martinez},
  {Martinovic}, {Martynov}, {Marx}, {Masalehdan}, {Mason}, {Massera},
  {Masserot}, {Massinger}, {Masso-Reid}, {Mastrogiovanni}, {Matas},
  {Mateu-Lucena}, {Matichard}, {Matiushechkina}, {Mavalvala}, {McCann},
  {McCarthy}, {McClelland}, {McClincy}, {McCormick}, {McCuller}, {McGhee},
  {McGuire}, {McIsaac}, {McIver}, {McRae}, {McWilliams}, {Meacher}, {Mehmet},
  {Mehta}, {Meijer}, {Melatos}, {Melchor}, {Mendell}, {Menendez-Vazquez},
  {Menoni}, {Mercer}, {Mereni}, {Merfeld}, {Merilh}, {Merritt}, {Merzougui},
  {Meshkov}, {Messenger}, {Messick}, {Meyers}, {Meylahn}, {Mhaske}, {Miani},
  {Miao}, {Michaloliakos}, {Michel}, {Michimura}, {Middleton}, {Milano},
  {Miller}, {Miller}, {Miller}, {Millhouse}, {Mills}, {Milotti}, {Minazzoli},
  {Minenkov}, {Mio}, {Mir}, {Miravet-Ten{\'e}s}, {Mishra}, {Mishra}, {Mistry},
  {Mitra}, {Mitrofanov}, {Mitselmakher}, {Mittleman}, {Miyakawa}, {Miyamoto},
  {Miyazaki}, {Miyo}, {Miyoki}, {Mo}, {Modafferi}, {Moguel}, {Mogushi},
  {Mohapatra}, {Mohite}, {Molina}, {Molina-Ruiz}, {Mondin}, {Montani}, {Moore},
  {Moraru}, {Morawski}, {More}, {Moreno}, {Moreno}, {Mori}, {Morisaki},
  {Moriwaki}, {Morr{\'a}s}, {Mours}, {Mow-Lowry}, {Mozzon}, {Muciaccia},
  {Mukherjee}, {Mukherjee}, {Mukherjee}, {Mukherjee}, {Mukherjee}, {Mukund},
  {Mullavey}, {Munch}, {Mu{\~n}iz}, {Murray}, {Musenich}, {Muusse}, {Nadji},
  {Nagano}, {Nagano}, {Nagar}, {Nakamura}, {Nakano}, {Nakano}, {Nakashima},
  {Nakayama}, {Napolano}, {Nardecchia}, {Narikawa}, {Naticchioni}, {Nayak},
  {Nayak}, {Negishi}, {Neil}, {Neilson}, {Nelemans}, {Nelson}, {Nery},
  {Neubauer}, {Neunzert}, {Ng}, {Ng}, {Nguyen}, {Nguyen}, {Nguyen}, {Nguyen
  Quynh}, {Ni}, {Nichols}, {Nishizawa}, {Nissanke}, {Nitoglia}, {Nocera},
  {Norman}, {North}, {Nozaki}, {Nu{\~n}o Siles}, {Nuttall}, {Oberling},
  {O'Brien}, {Obuchi}, {O'Dell}, {Oelker}, {Ogaki}, {Oganesyan}, {Oh}, {Oh},
  {Oh}, {Ohashi}, {Ohishi}, {Ohkawa}, {Ohme}, {Ohta}, {Okada}, {Okutani},
  {Okutomi}, {Olivetto}, {Oohara}, {Ooi}, {Oram}, {O'Reilly}, {Ormiston},
  {Ormsby}, {Ortega}, {O'Shaughnessy}, {O'Shea}, {Oshino}, {Ossokine},
  {Osthelder}, {Otabe}, {Ottaway}, {Overmier}, {Pace}, {Pagano}, {Page},
  {Pagliaroli}, {Pai}, {Pai}, {Palamos}, {Palashov}, {Palomba}, {Pan}, {Pan},
  {Panda}, {Pang}, {Pang}, {Pankow}, {Pannarale}, {Pant}, {Panther},
  {Paoletti}, {Paoli}, {Paolone}, {Parisi}, {Park}, {Park}, {Parker},
  {Pascucci}, {Pasqualetti}, {Passaquieti}, {Passuello}, {Patel}, {Pathak},
  {Patricelli}, {Patron}, {Paul}, {Payne}, {Pedraza}, {Pegoraro}, {Pele},
  {Pe{\~n}a Arellano}, {Penn}, {Perego}, {Pereira}, {Pereira}, {Perez},
  {P{\'e}rigois}, {Perkins}, {Perreca}, {Perri{\`e}s}, {Petermann},
  {Petterson}, {Pfeiffer}, {Pham}, {Phukon}, {Piccinni}, {Pichot},
  {Piendibene}, {Piergiovanni}, {Pierini}, {Pierro}, {Pillant}, {Pillas},
  {Pilo}, {Pinard}, {Pinto}, {Pinto}, {Piotrzkowski}, {Piotrzkowski},
  {Pirello}, {Pitkin}, {Placidi}, {Planas}, {Plastino}, {Pluchar}, {Poggiani},
  {Polini}, {Pong}, {Ponrathnam}, {Popolizio}, {Porter}, {Poulton}, {Powell},
  {Pracchia}, {Pradier}, {Prajapati}, {Prasai}, {Prasanna}, {Pratten},
  {Principe}, {Prodi}, {Prokhorov}, {Prosposito}, {Prudenzi}, {Puecher},
  {Punturo}, {Puosi}, {Puppo}, {P{\"u}rrer}, {Qi}, {Quetschke},
  {Quitzow-James}, {Qutob}, {Raab}, {Raaijmakers}, {Radkins}, {Radulesco},
  {Raffai}, {Rail}, {Raja}, {Rajan}, {Ramirez}, {Ramirez}, {Ramos-Buades},
  {Rana}, {Rapagnani}, {Rapol}, {Ray}, {Raymond}, {Raza}, {Razzano}, {Read},
  {Rees}, {Regimbau}, {Rei}, {Reid}, {Reid}, {Reitze}, {Relton}, {Renzini},
  {Rettegno}, {Reza}, {Rezac}, {Ricci}, {Richards}, {Richardson}, {Richardson},
  {Riemenschneider}, {Riles}, {Rinaldi}, {Rink}, {Rizzo}, {Robertson}, {Robie},
  {Robinet}, {Rocchi}, {Rodriguez}, {Rolland}, {Rollins}, {Romanelli},
  {Romano}, {Romel}, {Romero-Rodr{\'\i}guez}, {Romero-Shaw}, {Romie},
  {Ronchini}, {Rosa}, {Rose}, {Rosi{\'n}ska}, {Ross}, {Rowan}, {Rowlinson},
  {Roy}, {Roy}, {Roy}, {Rozza}, {Ruggi}, {Ruiz-Rocha}, {Ryan}, {Sachdev},
  {Sadecki}, {Sadiq}, {Sago}, {Saito}, {Saito}, {Sakai}, {Sakai},
  {Sakellariadou}, {Sakuno}, {Salafia}, {Salconi}, {Saleem}, {Salemi},
  {Samajdar}, {Sanchez}, {Sanchez}, {Sanchez}, {Sanchis-Gual}, {Sanders},
  {Sanuy}, {Saravanan}, {Sarin}, {Sassolas}, {Satari}, {Sathyaprakash}, {Sato},
  {Sato}, {Sauter}, {Savage}, {Sawada}, {Sawant}, {Sawant}, {Sayah},
  {Schaetzl}, {Scheel}, {Scheuer}, {Schiworski}, {Schmidt}, {Schmidt},
  {Schnabel}, {Schneewind}, {Schofield}, {Sch{\"o}nbeck}, {Schulte}, {Schutz},
  {Schwartz}, {Scott}, {Scott}, {Seglar-Arroyo}, {Sekiguchi}, {Sekiguchi},
  {Sellers}, {Sengupta}, {Sentenac}, {Seo}, {Sequino}, {Sergeev}, {Setyawati},
  {Shaffer}, {Shahriar}, {Shams}, {Shao}, {Sharma}, {Sharma}, {Shawhan},
  {Shcheblanov}, {Shibagaki}, {Shikauchi}, {Shimizu}, {Shimoda}, {Shimode},
  {Shinkai}, {Shishido}, {Shoda}, {Shoemaker}, {Shoemaker}, {ShyamSundar},
  {Sieniawska}, {Sigg}, {Singer}, {Singh}, {Singh}, {Singha}, {Sintes},
  {Sipala}, {Skliris}, {Slagmolen}, {Slaven-Blair}, {Smetana}, {Smith},
  {Smith}, {Soldateschi}, {Somala}, {Somiya}, {Son}, {Soni}, {Soni}, {Sordini},
  {Sorrentino}, {Sorrentino}, {Sotani}, {Soulard}, {Souradeep}, {Sowell},
  {Spagnuolo}, {Spencer}, {Spera}, {Srinivasan}, {Srivastava}, {Srivastava},
  {Staats}, {Stachie}, {Steer}, {Steinhoff}, {Steinlechner}, {Steinlechner},
  {Stevenson}, {Stops}, {Stover}, {Strain}, {Strang}, {Stratta}, {Strunk},
  {Sturani}, {Stuver}, {Sudhagar}, {Sudhir}, {Sugimoto}, {Suh}, {Sullivan},
  {Sullivan}, {Summerscales}, {Sun}, {Sun}, {Sunil}, {Sur}, {Suresh}, {Sutton},
  {Suzuki}, {Suzuki}, {Swinkels}, {Szczepa{\'n}czyk}, {Szewczyk}, {Tacca},
  {Tagoshi}, {Tait}, {Takahashi}, {Takahashi}, {Takamori}, {Takano}, {Takeda},
  {Takeda}, {Talbot}, {Talbot}, {Tanaka}, {Tanaka}, {Tanaka}, {Tanaka},
  {Tanaka}, {Tanasijczuk}, {Tanioka}, {Tanner}, {Tao}, {Tao}, {Tapia San
  Mart{\'\i}n}, {Taranto}, {Tasson}, {Telada}, {Tenorio}, {Terhune},
  {Terkowski}, {Thirugnanasambandam}, {Thomas}, {Thomas}, {Thomas}, {Thompson},
  {Thondapu}, {Thorne}, {Thrane}, {Tiwari}, {Tiwari}, {Tiwari}, {Toivonen},
  {Toland}, {Tolley}, {Tomaru}, {Tomigami}, {Tomura}, {Tonelli},
  {Torres-Forn{\'e}}, {Torrie}, {Tosta e Melo}, {T{\"o}yr{\"a}}, {Trapananti},
  {Travasso}, {Traylor}, {Trevor}, {Tringali}, {Tripathee}, {Troiano},
  {Trovato}, {Trozzo}, {Trudeau}, {Tsai}, {Tsai}, {Tsang}, {Tsang}, {Tsao},
  {Tse}, {Tso}, {Tsubono}, {Tsuchida}, {Tsukada}, {Tsuna}, {Tsutsui},
  {Tsuzuki}, {Turbang}, {Turconi}, {Tuyenbayev}, {Ubhi}, {Uchikata},
  {Uchiyama}, {Udall}, {Ueda}, {Uehara}, {Ueno}, {Ueshima}, {Unnikrishnan},
  {Uraguchi}, {Urban}, {Ushiba}, {Utina}, {Vahlbruch}, {Vajente}, {Vajpeyi},
  {Valdes}, {Valentini}, {Valsan}, {van Bakel}, {van Beuzekom}, {van den
  Brand}, {Van Den Broeck}, {Vander-Hyde}, {van der Schaaf}, {van Heijningen},
  {Vanosky}, {van Putten}, {van Remortel}, {Vardaro}, {Vargas}, {Varma},
  {Vas{\'u}th}, {Vecchio}, {Vedovato}, {Veitch}, {Veitch}, {Venneberg},
  {Venugopalan}, {Verkindt}, {Verma}, {Verma}, {Veske}, {Vetrano},
  {Vicer{\'e}}, {Vidyant}, {Viets}, {Vijaykumar}, {Villa-Ortega}, {Vinet},
  {Virtuoso}, {Vitale}, {Vo}, {Vocca}, {von Reis}, {von Wrangel}, {Vorvick},
  {Vyatchanin}, {Wade}, {Wade}, {Wagner}, {Walet}, {Walker}, {Wallace},
  {Wallace}, {Walsh}, {Wang}, {Wang}, {Wang}, {Ward}, {Warner}, {Was},
  {Washimi}, {Washington}, {Watchi}, {Weaver}, {Webster}, {Weinert},
  {Weinstein}, {Weiss}, {Weller}, {Weller}, {Wellmann}, {Wen}, {We{\ss}els},
  {Wette}, {Whelan}, {White}, {Whiting}, {Whittle}, {Wilken}, {Williams},
  {Williams}, {Williams}, {Williamson}, {Willis}, {Willke}, {Wilson},
  {Winkler}, {Wipf}, {Wlodarczyk}, {Woan}, {Woehler}, {Wofford}, {Wong}, {Wu},
  {Wu}, {Wu}, {Wu}, {Wysocki}, {Xiao}, {Xu}, {Yamada}, {Yamamoto}, {Yamamoto},
  {Yamamoto}, {Yamamoto}, {Yamashita}, {Yamazaki}, {Yang}, {Yang}, {Yang},
  {Yang}, {Yang}, {Yap}, {Yeeles}, {Yelikar}, {Ying}, {Yokogawa}, {Yokoyama},
  {Yokozawa}, {Yoo}, {Yoshioka}, {Yu}, {Yu}, {Yuzurihara}, {Zadro{\.z}ny},
  {Zanolin}, {Zeidler}, {Zelenova}, {Zendri}, {Zevin}, {Zhan}, {Zhang},
  {Zhang}, {Zhang}, {Zhang}, {Zhang}, {Zhao}, {Zhao}, {Zhao}, {Zhao}, {Zheng},
  {Zhou}, {Zhou}, {Zhu}, {Zhu}, {Zimmerman}, {Zlochower}, {Zucker}, \&
  {Zweizig}}]{ligo21}
{The LIGO Scientific Collaboration}, {the Virgo Collaboration}, {the KAGRA
  Collaboration}, {et~al.} 2021, arXiv e-prints, arXiv:2111.03606.
\newblock \doarXiv{2111.03606}

\bibitem[{{The LIGO Scientific Collaboration} {et~al.}(2023){The LIGO
  Scientific Collaboration}, {the Virgo Collaboration}, {the KAGRA
  Collaboration}, {Abbott}, {Abe}, {Acernese}, {Ackley}, {Adhicary},
  {Adhikari}, {Adhikari}, {Adkins}, {Adya}, {Affeldt}, {Agarwal}, {Agathos},
  {Aguiar}, {Aiello}, {Ain}, {Ajith}, {Akutsu}, {Albanesi}, {Alfaidi},
  {Al-Jodah}, {All{\'e}n{\'e}}, {Allocca}, {Almualla}, {Altin}, {Amato},
  {Amez-Droz}, {Amorosi}, {Anand}, {Ananyeva}, {Andersen}, {Anderson},
  {Anderson}, {Andia}, {Ando}, {Andrade}, {Andres}, {Andr{\'e}s-Carcasona},
  {Andri{\'c}}, {Ansoldi}, {Antelis}, {Antier}, {Aoumi}, {Apostolatos},
  {Appavuravther}, {Appert}, {Apple}, {Arai}, {Araya}, {Araya}, {Areeda},
  {Ar{\`e}ne}, {Aritomi}, {Arnaud}, {Arogeti}, {Aronson}, {Arun}, {Asada},
  {Ashton}, {Aso}, {Assiduo}, {Assis de Souza Melo}, {Aston}, {Astone},
  {Aubin}, {AultONeal}, {Babak}, {Badalyan}, {Badaracco}, {Badger}, {Bae},
  {Bagnasco}, {Bai}, {Baier}, {Baiotti}, {Baird}, {Bajpai}, {Baka}, {Ball},
  {Ballardin}, {Ballmer}, {Baltus}, {Banagiri}, {Banerjee}, {Bankar}, {Baral},
  {Barayoga}, {Barber}, {Barish}, {Barker}, {Barneo}, {Barone}, {Barr},
  {Barsotti}, {Barsuglia}, {Barta}, {Barthelmy}, {Barton}, {Bartos}, {Basak},
  {Basalaev}, {Bassiri}, {Basti}, {Bawaj}, {Bayley}, {Baylor}, {Bazzan},
  {B{\'e}csy}, {Bedakihale}, {Beirnaert}, {Bejger}, {Bell}, {Benedetto},
  {Beniwal}, {Benoit}, {Bentley}, {Ben Yaala}, {Bera}, {Berbel}, {Bergamin},
  {Berger}, {Bernuzzi}, {Beroiz}, {Berry}, {Bersanetti}, {Bertolini},
  {Betzwieser}, {Beveridge}, {Bevins}, {Bhandare}, {Bhandari}, {Bhardwaj},
  {Bhatt}, {Bhattacharjee}, {Bhaumik}, {Bianchi}, {Bilenko}, {Bilicki},
  {Billingsley}, {Bini}, {Birnholtz}, {Biscans}, {Bischi}, {Biscoveanu},
  {Bisht}, {Biswas}, {Bitossi}, {Bizouard}, {Blackburn}, {Blair}, {Blair},
  {Blair}, {Bobba}, {Bode}, {Bo{\"e}r}, {Bogaert}, {Boileau}, {Boldrini},
  {Bolingbroke}, {Bonavena}, {Bondarescu}, {Bondu}, {Bonilla}, {Bonilla},
  {Bonnand}, {Booker}, {Bork}, {Boschi}, {Bose}, {Bose}, {Bossilkov},
  {Boudart}, {Bouffanais}, {Bozzi}, {Bradaschia}, {Brady}, {Braglia}, {Branch},
  {Branchesi}, {Brau}, {Breschi}, {Briant}, {Brillet}, {Brinkmann}, {Brockill},
  {Brooks}, {Brooks}, {Brown}, {Brunett}, {Bruno}, {Bruntz}, {Bryant}, {Bucci},
  {Buchanan}, {Bulashenko}, {Bulik}, {Bulten}, {Buonanno}, {Burtnyk},
  {Buscicchio}, {Buskulic}, {Buy}, {Byer}, {Cabourn Davies}, {Cabras},
  {Cabrita}, {Cadonati}, {Caesar}, {Cagnoli}, {Cahillane}, {Calder{\'o}n
  Bustillo}, {Callaghan}, {Callister}, {Calloni}, {Camp}, {Canepa}, {Caneva
  Santoro}, {Cannavacciuolo}, {Cannon}, {Cao}, {Cao}, {Capistran}, {Capocasa},
  {Capote}, {Carapella}, {Carbognani}, {Carlassara}, {Carlin}, {Carpinelli},
  {Carter}, {Carullo}, {Casanueva Diaz}, {Casentini}, {Castaldi},
  {Castro-Lucas}, {Caudill}, {Cavagli{\`a}}, {Cavalieri}, {Cella},
  {Cerd{\'a}-Dur{\'a}n}, {Cesarini}, {Chaibi}, {Chakalis}, {Chalathadka
  Subrahmanya}, {Champion}, {Chan}, {Chan}, {Chandra}, {Chang}, {Chang},
  {Chanial}, {Chao}, {Chapman-Bird}, {Charlton}, {Charlton},
  {Chassande-Mottin}, {Chastain}, {Chatterjee}, {Chatterjee}, {Chatterjee},
  {Chaturvedi}, {Chaty}, {Chatziioannou}, {Chen}, {Chen}, {Chen}, {Chen},
  {Chen}, {Chen}, {Chen}, {Chen}, {Cheng}, {Chessa}, {Cheung}, {Chia},
  {Chiadini}, {Chiang}, {Chiang}, {Chiarini}, {Chiba}, {Chiba}, {Chierici},
  {Chincarini}, {Chiofalo}, {Chiummo}, {Choudhary}, {Christensen}, {Chua},
  {Chung}, {Ciani}, {Ciecielag}, {Cie{\'s}lar}, {Cifaldi}, {Ciobanu}, {Ciolfi},
  {Clara}, {Clark}, {Clarke}, {Clearwater}, {Clesse}, {Cleva}, {Coccia},
  {Codazzo}, {Cohadon}, {Colleoni}, {Collette}, {Colombo}, {Colpi}, {Compton},
  {Conti}, {Cooper}, {Corban}, {Corbitt}, {Cordero-Carri{\'o}n}, {Corezzi},
  {Cornish}, {Corsi}, {Cortese}, {Coschizza}, {Cottingham}, {Coughlin},
  {Coulon}, {Countryman}, {Coupechoux}, {Cousins}, {Couvares}, {Coward},
  {Cowart}, {Cowburn}, {Coyne}, {Coyne}, {Craig}, {Creighton}, {Creighton},
  {Criswell}, {Crockett-Gray}, {Croquette}, {Crowder}, {Cudell}, {Cullen},
  {Cumming}, {Cummings}, {Cuoco}, {Cury{\l}o}, {Dabadie}, {Dal Canton},
  {Dall'Osso}, {D{\'a}lya}, {D'Angelo}, {Danilishin}, {D'Antonio}, {Danzmann},
  {Darroch}, {Darsow-Fromm}, {Dasgupta}, {Datrier}, {Datta}, {Dattilo}, {Dave},
  {Davenport}, {Davier}, {Davis}, {Davis}, {Daw}, {Dax}, {DeBra},
  {Deenadayalan}, {Degallaix}, {De Laurentis}, {Del{\'e}glise}, {Del Favero},
  {De Lillo}, {De Lillo}, {Dell'Aquila}, {Del Pozzo}, {De Matteis}, {D'Emilio},
  {Demos}, {Dent}, {Depasse}, {De Pietri}, {De Rosa}, {De Rossi}, {DeSalvo},
  {De Simone}, {Dhurandhar}, {Diab}, {Diamond}, {D{\'\i}az}, {Didio},
  {Dietrich}, {Di Fiore}, {Di Fronzo}, {Di Giorgio}, {Di Giovanni}, {Di
  Giovanni}, {Di Girolamo}, {Diksha}, {Di Lieto}, {Di Michele}, {Di Pace}, {Di
  Palma}, {Di Renzo}, {Divyajyoti}, {Dmitriev}, {Doctor}, {Dohmen}, {Doleva},
  {Donahue}, {D'Onofrio}, {Donovan}, {Dooley}, {Dooney}, {Doravari}, {Dorosh},
  {Drago}, {Driggers}, {Drori}, {Ducoin}, {Dunn}, {Dupletsa}, {Durante},
  {D'Urso}, {Duverne}, {Dwyer}, {Eassa}, {Easter}, {Ebersold}, {Eckhardt},
  {Eddolls}, {Edelman}, {Edo}, {Edy}, {Effler}, {Eichholz}, {Eisenmann},
  {Eisenstein}, {Ejlli}, {Engelby}, {Engl}, {Errico}, {Essick}, {Estell{\'e}s},
  {Estevez}, {Etzel}, {Evans}, {Evans}, {Evans}, {Evstafyeva}, {Ewing},
  {Fabrizi}, {Faedi}, {Fafone}, {Fair}, {Fairhurst}, {Fan}, {Fan}, {Farah},
  {Farr}, {Farr}, {Fauchon-Jones}, {Favaro}, {Favata}, {Fays}, {Feicht},
  {Fejer}, {Fenyvesi}, {Ferguson}, {Fernandez-Galiana}, {Ferrante}, {Ferreira},
  {Fidecaro}, {Figura}, {Fiori}, {Fiori}, {Fishbach}, {Fisher}, {Fittipaldi},
  {Fiumara}, {Flaminio}, {Fleischer}, {Fleming}, {Floden}, {Fong}, {Font},
  {Fornal}, {Forsyth}, {Franke}, {Frasca}, {Frasconi}, {Freed}, {Frei},
  {Freise}, {Freitas}, {Frey}, {Fritschel}, {Frolov}, {Fronz{\'e}}, {Fujimoto},
  {Fukunaga}, {Fulda}, {Fyffe}, {Gabbard}, {Gabella}, {Gadre}, {Gaglani},
  {Gair}, {Gais}, {Galaudage}, {Gallardo}, {Gamba}, {Ganapathy}, {Ganguly},
  {Gao}, {Gaonkar}, {Garaventa}, {Garcia-Bellido}, {Garc{\'\i}a-N{\'u}{\~n}ez},
  {Garc{\'\i}a-Quir{\'o}s}, {Gardner}, {Gargiulo}, {Garufi}, {Gasbarra},
  {Gateley}, {Gayathri}, {Gemme}, {Gennai}, {George}, {Gerberding}, {Gergely},
  {Ghonge}, {Ghosh}, {Ghosh}, {Ghosh}, {Ghosh}, {Ghosh}, {Giacoppo}, {Giaime},
  {Giardina}, {Gibson}, {Gier}, {Giri}, {Gissi}, {Gkaitatzis}, {Glanzer},
  {Gleckl}, {Glotin}, {Godfrey}, {Godwin}, {Goetz}, {Goetz}, {Golomb},
  {Goncharov}, {Gonz{\'a}lez}, {Gosselin}, {Gouaty}, {Gould}, {Goyal}, {Grace},
  {Grado}, {Graham}, {Granata}, {Granata}, {Gras}, {Grassia}, {Gray}, {Gray},
  {Greco}, {Green}, {Green}, {Green}, {Green}, {Gretarsson}, {Gretarsson},
  {Griffith}, {Griffiths}, {Griggs}, {Grignani}, {Grimaldi}, {Grote}, {Gruson},
  {Guerra}, {Guetta}, {Guidi}, {Guimaraes}, {Gulati}, {Gulminelli}, {Gunny},
  {Guo}, {Guo}, {Gupta}, {Gupta}, {Gupta}, {Gupta}, {Gupta}, {Gupta}, {Gurs},
  {Gushima}, {Gustafson}, {Gutierrez}, {Guzman}, {Haegel}, {Hain}, {Haino},
  {Halim}, {Hall}, {Hamilton}, {Hammond}, {Han}, {Haney}, {Hanks}, {Hanna},
  {Hannam}, {Hannuksela}, {Hansen}, {Hanson}, {Harada}, {Harder}, {Haris},
  {Harmark}, {Harms}, {Harry}, {Harry}, {Hartwig}, {Haskell}, {Haster},
  {Hathaway}, {Haughian}, {Hayakawa}, {Hayama}, {Hayes}, {Healy}, {Heffernan},
  {Heidmann}, {Heintze}, {Heinze}, {Heinzel}, {Heitmann}, {Hellman}, {Hello},
  {Helmling-Cornell}, {Hemming}, {Hendry}, {Heng}, {Hennes}, {Hennig},
  {Hennig}, {Henshaw}, {Hernandez Vivanco}, {Heurs}, {Hewitt}, {Higginbotham},
  {Hild}, {Hill}, {Himemoto}, {Hines}, {Hirata}, {Hirose}, {Ho}, {Hochheim},
  {Hofman}, {Hohmann}, {Holcomb}, {Holland}, {Holley-Bockelmann}, {Hollows},
  {Holmes}, {Holt}, {Holz}, {Hong}, {Hornung}, {Hoshino}, {Hough}, {Hourihane},
  {Howell}, {Howell}, {Hoy}, {Hoyland}, {Hsieh}, {Hsieh}, {Hsiung}, {Hsu},
  {Hu}, {Hu}, {Huang}, {Huang}, {Huang}, {Huang}, {H{\"u}bner}, {Huddart},
  {Hughey}, {Hui}, {Hui}, {Husa}, {Huttner}, {Huxford}, {Huynh-Dinh}, {Hyland},
  {Iakovlev}, {Iandolo}, {Idzkowski}, {Iess}, {Inayoshi}, {Inoue}, {Iorio},
  {Iosif}, {Irwin}, {Isi}, {Ismail}, {Itoh}, {Iyer}, {JaberianHamedan},
  {Jacqmin}, {Jacquet}, {Jadhav}, {Jadhav}, {Jain}, {Jain}, {James}, {Jan},
  {Jani}, {Janiurek}, {Janquart}, {Janssens}, {Janthalur}, {Jaraba},
  {Jaranowski}, {Jarov}, {Jasal}, {Jaume}, {Javed}, {Jenkins}, {Jenner},
  {Jennings}, {Jia}, {Jiang}, {Liu}, {Jin}, {Johansmeyer}, {Johns}, {Johnson},
  {Johnston}, {Johny}, {Jones}, {Jones}, {Jones}, {Jones}, {Jones}, {Joshi},
  {Ju}, {Jung}, {Junker}, {Juste}, {Kajita}, {Kalaghatgi}, {Kalogera}, {Kamai},
  {Kamiizumi}, {Kanda}, {Kandhasamy}, {Kang}, {Kanner}, {Kapadia}, {Kapasi},
  {Karat}, {Karathanasis}, {Karki}, {Kasamatsu}, {Kas-danouche}, {Kashyap},
  {Kasprzack}, {Kastaun}, {Kato}, {Katsanevas}, {Katsavounidis}, {Katsuren},
  {Katzman}, {Kaur}, {Kawabe}, {Kawazoe}, {K{\'e}f{\'e}lian}, {Keitel},
  {Kellard}, {Kelley-Derzon}, {Kennington}, {Key}, {Khadka}, {Khalili}, {Khan},
  {Khanam}, {Khazanov}, {Khursheed}, {Kijbunchoo}, {Kim}, {Kim}, {Kim}, {Kim},
  {Kim}, {Kim}, {Kim}, {Kim}, {Kimball}, {Kimura}, {Kinley-Hanlon},
  {Kirchhoff}, {Kissel}, {Kiyota}, {Klimenko}, {Klinger}, {Knee}, {Knust},
  {Kobayashi}, {Koch}, {Koehlenbeck}, {Koekoek}, {Kohri}, {Kokeyama}, {Koley},
  {Koliadko}, {Kolitsidou}, {Kolstein}, {Kondrashov}, {Kong}, {Kontos},
  {Korobko}, {Kossak}, {Kouvatsos}, {Kovalam}, {Koyama}, {Kozak}, {Kranzhoff},
  {Kranzhoff}, {Kringel}, {Krishnendu}, {Kr{\'o}lak}, {Kuehn}, {Kuijer},
  {Kukihara}, {Kulkarni}, {Kumar}, {Kumar}, {Kumar}, {Kumar}, {Kumar}, {Kume},
  {Kuns}, {Kuroyanagi}, {Kuwahara}, {Kwak}, {Lacaille}, {Lagabbe}, {Laghi},
  {Lakkis}, {Lalande}, {Lalleman}, {Lamberts}, {Landry}, {Lane}, {Lang},
  {Lange}, {Lantz}, {La Rana}, {La Rosa}, {Lartaux-Vollard}, {Lasky},
  {Lawrence}, {Laxen}, {Lazzarini}, {Lazzaro}, {Leaci}, {Leavey}, {LeBohec},
  {Lecoeuche}, {Lee}, {Lee}, {Lee}, {Lee}, {Lee}, {Lee}, {Lee}, {Legred},
  {Lehmann}, {Lehner}, {Lema{\^\i}tre}, {Lenti}, {Leonardi}, {Leonova},
  {Leroy}, {Letendre}, {Lethuillier}, {Levesque}, {Levin}, {Leyde}, {Li}, {Li},
  {Li}, {Li}, {Lin}, {Lin}, {Lin}, {Lin}, {Lin}, {Lin}, {Lin}, {Lin}, {Linde},
  {Linker}, {Littenberg}, {Liu}, {Liu}, {Llamas}, {Lo}, {Lo}, {London},
  {Longo}, {Lopez}, {Lopez Portilla}, {Lorenzini}, {Loriette}, {Lormand},
  {Losurdo}, {Lott}, {Lough}, {Loughlin}, {Lousto}, {Lovelace}, {Lowry},
  {L{\"u}ck}, {Lumaca}, {Lundgren}, {Lung}, {Lussier}, {Lynam}, {Ma}, {Ma},
  {Ma'arif}, {Macas}, {MacInnis}, {Macleod}, {MacMillan}, {Macquet},
  {Maga{\~n}a Hernandez}, {Magazz{\`u}}, {Magee}, {Maggiore}, {Magnozzi},
  {Mahesh}, {Mahesh}, {Maini}, {Majorana}, {Makarem}, {Maliakal}, {Malik},
  {Man}, {Mandic}, {Mangano}, {Mannix}, {Mansell}, {Mansingh}, {Manske},
  {Mantovani}, {Mapelli}, {Marchesoni}, {Mar{\'\i}n Pina}, {Marion},
  {M{\'a}rka}, {M{\'a}rka}, {Markakis}, {Markosyan}, {Markowitz}, {Maros},
  {Marquina}, {Marsat}, {Martelli}, {Martin}, {Martin}, {Martinez}, {Martinez},
  {Martinez}, {Martinez}, {Martinovic}, {Martynov}, {Marx}, {Masalehdan},
  {Mason}, {Masserot}, {Masso Reid}, {Mastrodicasa}, {Mastrogiovanni},
  {Mateu-Lucena}, {Matiushechkina}, {Matsunaga}, {Mavalvala}, {McCarthy},
  {McClelland}, {McClincy}, {McCormick}, {McCuller}, {McGhee}, {McGinn},
  {McIsaac}, {McIver}, {McLeod}, {McRae}, {McWilliams}, {Meacher}, {Mehmet},
  {Mehta}, {Meijer}, {Melatos}, {Mendell}, {Menendez-Vazquez}, {Menoni},
  {Mercer}, {Mereni}, {Merfeld}, {Merilh}, {Merritt}, {Merzougui}, {Messenger},
  {Messick}, {Meyers}, {Meylahn}, {Mhaske}, {Miani}, {Miao}, {Michaloliakos},
  {Michel}, {Michimura}, {Middleton}, {Mihaylov}, {Miller}, {Miller}, {Miller},
  {Miller}, {Millhouse}, {Mills}, {Milotti}, {Minenkov}, {Mio}, {Mir},
  {Miravet-Ten{\'e}s}, {Mishra}, {Mishra}, {Mishra}, {Mistry}, {Mitchell},
  {Mitra}, {Mitrofanov}, {Mitselmakher}, {Mittleman}, {Miyakawa}, {Miyoki},
  {Mo}, {Modafferi}, {Moguel}, {Mohapatra}, {Mohite}, {Molina-Ruiz}, {Mondal},
  {Mondin}, {Montani}, {Moore}, {Moragues}, {Moraru}, {Morawski}, {More},
  {More}, {Moreno}, {Moreno}, {Morisaki}, {Moriwaki}, {Morras}, {Moscatello},
  {Mours}, {Mow-Lowry}, {Mozzon}, {Muciaccia}, {Mukherjee}, {Mukherjee},
  {Mukherjee}, {Mukherjee}, {Mukund}, {Mullavey}, {Munch}, {Mu{\~n}iz},
  {Murray}, {Murray-Dean}, {Muusse}, {Nadji}, {Nagar}, {Nagar}, {Nagarajan},
  {Nakamura}, {Nakano}, {Nakano}, {Nakayama}, {Napolano}, {Nardecchia},
  {Narikawa}, {Narola}, {Naticchioni}, {Nayak}, {Neil}, {Neilson}, {Nelson},
  {Nelson}, {Nery}, {Nesseris}, {Neunzert}, {Ng}, {Ng}, {Nguyen}, {Nguyen},
  {Nguyen}, {Nguyen}, {Nguyen Quynh}, {Nichols}, {Nieradka}, {Nishino},
  {Nishizawa}, {Nissanke}, {Nitoglia}, {Niu}, {Nocera}, {Norman}, {North},
  {Novak}, {Nu{\~n}o Siles}, {Nurbek}, {Nuttall}, {Oberling}, {O'Dell},
  {Oelker}, {Oertel}, {Oganesyan}, {Oh}, {Oh}, {Oh}, {O'Hanlon}, {Ohashi},
  {Ohashi}, {Ohkawa}, {Ohme}, {Ohta}, {Oliveira}, {Oliveri}, {Oohara},
  {O'Reilly}, {Ormiston}, {Ormsby}, {Orselli}, {O'Shaughnessy}, {O'Shea},
  {Oshima}, {Oshino}, {Ossokine}, {Osthelder}, {Ottaway}, {Overmier}, {Pace},
  {Pagano}, {Page}, {Pai}, {Pai}, {Pal}, {Palashov}, {P{\'a}lfi}, {Palomba},
  {Pan}, {Panda}, {Pang}, {Pannarale}, {Pant}, {Panther}, {Paoletti}, {Paoli},
  {Paolone}, {Papalexakis}, {Pappas}, {Parisi}, {Park}, {Parker}, {Pascucci},
  {Pasqualetti}, {Passaquieti}, {Passuello}, {Patel}, {Pathak}, {Patra},
  {Patricelli}, {Patron}, {Paul}, {Payne}, {Pearce}, {Pedraza}, {Pedurand},
  {Pegna}, {Pegoraro}, {Pele}, {Pe{\~n}a Arellano}, {Penn}, {Perego},
  {Pereira}, {Perez}, {P{\'e}rigois}, {Perkins}, {Perreca}, {Perri{\`e}s},
  {Perry}, {Pesios}, {Petermann}, {Petrillo}, {Pfeiffer}, {Pham}, {Pham},
  {Phukon}, {Phurailatpam}, {Piccinni}, {Pichot}, {Piendibene}, {Piergiovanni},
  {Pierini}, {Pierra}, {Pierro}, {Pillant}, {Pillas}, {Pilo}, {Pinard},
  {Pineda-Bosque}, {Pinto}, {Piotrzkowski}, {Piotrzkowski}, {Pirello},
  {Pitkin}, {Placidi}, {Placidi}, {Planas}, {Plastino}, {Poggiani}, {Polini},
  {Pompili}, {Pong}, {Ponrathnam}, {Porcelli}, {Portell}, {Porter},
  {Posnansky}, {Poulton}, {Powell}, {Powell}, {Pracchia}, {Pradier},
  {Prajapati}, {Prasai}, {Prasanna}, {Pratten}, {Principe}, {Prodi},
  {Prokhorov}, {Prosposito}, {Prudenzi}, {Puecher}, {Pullin}, {Punturo},
  {Puosi}, {Puppo}, {P{\"u}rrer}, {Qi}, {Quetschke}, {Quinonez},
  {Quitzow-James}, {Raab}, {Raaijmakers}, {Radulesco}, {Raffai}, {Rail},
  {Raja}, {Rajan}, {Ramirez}, {Ramirez}, {Ramos-Buades}, {Rana}, {Rana},
  {Randel}, {Rangnekar}, {Rapagnani}, {Ray}, {Raymond}, {Raza}, {Razzano},
  {Read}, {Regimbau}, {Rei}, {Reid}, {Reid}, {Reitze}, {Relton}, {Renzini},
  {Rettegno}, {Revenu}, {Reza}, {Rezac}, {Rezaei}, {Ricci}, {Richards},
  {Richardson}, {Rijal}, {Riles}, {Riley}, {Rinaldi}, {Robertson}, {Robertson},
  {Robinet}, {Rocchi}, {Rodriguez}, {Rolland}, {Rollins}, {Romanelli},
  {Romano}, {Romel}, {Romero}, {Romero-Shaw}, {Romie}, {Ronchini}, {Roocke},
  {Rosa}, {Rosauer}, {Rose}, {Rosi{\'n}ska}, {Ross}, {Rossello}, {Roussel},
  {Rowan}, {Rowlinson}, {Roy}, {Royzman}, {Rozza}, {Ruggi}, {Ruiz Morales},
  {Ruiz-Rocha}, {Ryan}, {Sachdev}, {Sadecki}, {Sadiq}, {Saffarieh}, {Saha},
  {Saha}, {Saito}, {Sakai}, {Sakellariadou}, {Sako}, {Sakon}, {Salafia},
  {Salces-Carcoba}, {Salconi}, {Saleem}, {Salemi}, {Sall{\'e}}, {Samajdar},
  {Sanchez}, {Sanchez}, {Sanchez}, {Sanchis-Gual}, {Sanders}, {Sanuy},
  {Saravanan}, {Sarin}, {Sasli}, {Sassi}, {Sassolas}, {Satari}, {Sauter},
  {Savage}, {Savant}, {Sawada}, {Sawant}, {Sayah}, {Schaetzl}, {Scheel},
  {Scherf}, {Scheuer}, {Schiworski}, {Schmidt}, {Schmidt}, {Schmitz},
  {Schnabel}, {Schneewind}, {Schofield}, {Sch{\"o}nbeck}, {Schuler}, {Schulte},
  {Schutz}, {Schwartz}, {Scott}, {Scott}, {Seetharamu}, {Seglar-Arroyo},
  {Sekiguchi}, {Sellers}, {Sengupta}, {Sentenac}, {Seo}, {Sequino}, {Sergeev},
  {Servignat}, {Setyawati}, {Shaffer}, {Shahriar}, {Shaikh}, {Shams}, {Shao},
  {Sharma}, {Sharma Chaudhary}, {Shawhan}, {Shcheblanov}, {Sheela}, {Shen},
  {Shepard}, {Sheridan}, {Shikano}, {Shikauchi}, {Shimizu}, {Shimode},
  {Shinkai}, {Shoemaker}, {Shoemaker}, {ShyamSundar}, {Sider}, {Siegel},
  {Sieniawska}, {Sigg}, {Silenzi}, {Singer}, {Singh}, {Singh}, {Singh},
  {Singha}, {Sintes}, {Sipala}, {Skliris}, {Slagmolen}, {Slaven-Blair},
  {Smetana}, {Smith}, {Smith}, {Smith}, {Soldateschi}, {Somala}, {Somiya},
  {Soni}, {Soni}, {Sordini}, {Sorrentino}, {Sorrentino}, {Sotani}, {Soulard},
  {Souradeep}, {Sowell}, {Spagnuolo}, {Spencer}, {Spera}, {Spinicelli},
  {Srivastava}, {Srivastava}, {Stachie}, {Stachurski}, {Steer}, {Steinlechner},
  {Steinlechner}, {Stergioulas}, {StPierre}, {Strang}, {Stratta}, {Strong},
  {Strunk}, {Sturani}, {Stuver}, {Suchenek}, {Sudhagar}, {Sueltmann},
  {Sugiyama}, {Suh}, {Sullivan}, {Summerscales}, {Sun}, {Sunil}, {Sur},
  {Suresh}, {Sutton}, {Suzuki}, {Suzuki}, {Swinkels}, {Syx},
  {Szczepa{\'n}czyk}, {Szewczyk}, {Tacca}, {Tagoshi}, {Tait}, {Takahashi},
  {Takahashi}, {Takamori}, {Takano}, {Takeda}, {Takeda}, {Talbot}, {Talbot},
  {Tamaki}, {Tamanini}, {Tanabe}, {Tanaka}, {Tanaka}, {Tanasijczuk}, {Tanioka},
  {Tanner}, {Tao}, {Tao}, {Tapia}, {Tapia San Mart{\'\i}n}, {Tarafder},
  {Taranto}, {Taruya}, {Tasson}, {Teloi}, {Tenorio}, {Terhune}, {Terkowski},
  {Themann}, {Thirugnanasambandam}, {Thomas}, {Thomas}, {Thomas}, {Thomas},
  {Thompson}, {Thondapu}, {Thorne}, {Thrane}, {Tiwari}, {Tiwari}, {Tiwari},
  {Toivonen}, {Tolley}, {Tomaru}, {Tomita}, {Tomura}, {Tonelli},
  {Torres-Forn{\'e}}, {Torrie}, {Tosta e Melo}, {Tournefier}, {Trapananti},
  {Travasso}, {Traylor}, {Trenado}, {Trevor}, {Tringali}, {Tripathee},
  {Troiano}, {Trovato}, {Trozzo}, {Trudeau}, {Tsang}, {Tsang}, {Tse}, {Tso},
  {Tsuchida}, {Tsukada}, {Tsutsui}, {Turbang}, {Turconi}, {Turski},
  {Tuyenbayev}, {Ubach}, {Ubhi}, {Uchikata}, {Uchiyama}, {Udall}, {Uehara},
  {Ueno}, {Unnikrishnan}, {Ushiba}, {Utina}, {Vahlbruch}, {Vaidya}, {Vajente},
  {Vajpeyi}, {Valdes}, {Valentini}, {Vallero}, {Valsan}, {van Bakel}, {van
  Beuzekom}, {van Dael}, {van den Brand}, {Van Den Broeck}, {Vander-Hyde}, {van
  der Sluys}, {Van de Walle}, {van Dongen}, {van Haevermaet}, {van Heijningen},
  {Vanosky}, {van Putten}, {van Ranst}, {van Remortel}, {Vardaro}, {Vargas},
  {Varma}, {Vas{\'u}th}, {Vecchio}, {Vedovato}, {Veitch}, {Veitch},
  {Venneberg}, {Venugopalan}, {Verdier}, {Verkindt}, {Verma}, {Verma},
  {Vermeulen}, {Veske}, {Vetrano}, {Vicer{\'e}}, {Vidyant}, {Viets},
  {Vijaykumar}, {Villa-Ortega}, {Vina}, {Vincent}, {Vinet}, {Viret},
  {Virtuoso}, {Vitale}, {Vocca}, {Voigt}, {von Reis}, {von Wrangel}, {Vorvick},
  {Vyatchanin}, {Wade}, {Wade}, {Wagner}, {Walet}, {Walker}, {Wallace},
  {Wallace}, {Wang}, {Wang}, {Wang}, {Ward}, {Warner}, {Was}, {Washimi},
  {Washington}, {Watada}, {Watarai}, {Watchi}, {Wayt}, {Weaver}, {Weaving},
  {Webster}, {Weinert}, {Weinstein}, {Weiss}, {Weller}, {Weller}, {Wellmann},
  {Wen}, {We{\ss}els}, {Wette}, {Whelan}, {White}, {Whiting}, {Whittle},
  {Wilk}, {Wilken}, {Willetts}, {Williams}, {Williams}, {Williamson}, {Willis},
  {Willke}, {Wipf}, {Woan}, {Woehler}, {Wofford}, {Wong}, {Wong}, {Wong},
  {Wright}, {Wu}, {Wu}, {Wu}, {Wysocki}, {Xiao}, {Xu}, {Yadav}, {Yamada},
  {Yamamoto}, {Yamamoto}, {Yamamoto}, {Yamamoto}, {Yamamoto}, {Yamashita},
  {Yamazaki}, {Yang}, {Yang}, {Yang}, {Yap}, {Yeeles}, {Yelikar}, {Yeung},
  {Yokoyama}, {Yokozawa}, {Yoo}, {Yu}, {Yu}, {Yuzurihara}, {Zadro{\.z}ny},
  {Zannelli}, {Zanolin}, {Zeeshan}, {Zeidler}, {Zelenova}, {Zendri}, {Zevin},
  {Zhang}, {Zhang}, {Zhang}, {Zhang}, {Zhang}, {Zhao}, {Zhao}, {Zhao}, {Zheng},
  {Zhong}, {Zhou}, {Zhu}, {Zhu}, {Zimmerman}, {Zucker}, \& {Zweizig}}]{ligo23}
---. 2023, arXiv e-prints, arXiv:2302.03676, \dodoi{10.48550/arXiv.2302.03676}

\bibitem[{{Tichy} \& {Marronetti}(2008)}]{tichy08}
{Tichy}, W., \& {Marronetti}, P. 2008, \prd, 78, 081501,
  \dodoi{10.1103/PhysRevD.78.081501}

\bibitem[{{Tong} {et~al.}(2022){Tong}, {Galaudage}, \& {Thrane}}]{tong22}
{Tong}, H., {Galaudage}, S., \& {Thrane}, E. 2022, \prd, 106, 103019,
  \dodoi{10.1103/PhysRevD.106.103019}

\bibitem[{{Tutukov} \& {Yungelson}(1993)}]{tutukov93}
{Tutukov}, A.~V., \& {Yungelson}, L.~R. 1993, \mnras, 260, 675,
  \dodoi{10.1093/mnras/260.3.675}

\bibitem[{{van den Heuvel}(1976)}]{vandenheuvel76}
{van den Heuvel}, E.~P.~J. 1976, in Structure and Evolution of Close Binary
  Systems, ed. P.~{Eggleton}, S.~{Mitton}, \& J.~{Whelan}, Vol.~73, 35

\bibitem[{{van den Heuvel} {et~al.}(2017){van den Heuvel}, {Portegies Zwart},
  \& {de Mink}}]{vandenheuvel17}
{van den Heuvel}, E.~P.~J., {Portegies Zwart}, S.~F., \& {de Mink}, S.~E. 2017,
  \mnras, 471, 4256, \dodoi{10.1093/mnras/stx1430}

\bibitem[{{Woosley}(2017)}]{woosley17}
{Woosley}, S.~E. 2017, \apj, 836, 244, \dodoi{10.3847/1538-4357/836/2/244}

\bibitem[{{Woosley} {et~al.}(2007){Woosley}, {Blinnikov}, \&
  {Heger}}]{woosley07}
{Woosley}, S.~E., {Blinnikov}, S., \& {Heger}, A. 2007, \nat, 450, 390,
  \dodoi{10.1038/nature06333}

\end{thebibliography}
\bibliographystyle{aasjournal}



\end{document}